%% file: paper.tex
\documentclass[useAMS,usenatbib]{mn2e}
\usepackage{graphicx}
\usepackage{subfigure}
\usepackage{amsmath}
\usepackage{amssymb}
\usepackage{comment}
\usepackage{color}

\def\ltsima{$\; \buildrel < \over \sim \;$}
\def\simlt{\lower.5ex\hbox{\ltsima}}
\def\gtsima{$\; \buildrel > \over \sim \;$}
\def\simgt{\lower.5ex\hbox{\gtsima}}

\title[]{The formation of CDM haloes II: collapse time and tides}
\author[Borzyszkowski, Ludlow \& Porciani]{Mikolaj Borzyszkowski\thanks{E-mail: mikolajb@astro.uni-bonn.de}, Aaron D. Ludlow and Cristiano Porciani\\
Argelander-Institut f\"ur Astronomie, Auf dem H\"ugel 71, D-53121 Bonn, Germany}
\begin{document}

\include{./adsmacro}

\date{\today}

\pagerange{\pageref{firstpage}--\pageref{lastpage}} \pubyear{2014}

\maketitle

\label{firstpage}

\begin{abstract}
We use two cosmological simulations of structure formation in the $\Lambda$CDM scenario to study 
the evolutionary histories of dark-matter haloes and to characterize the Lagrangian regions from 
which they form. We focus on haloes identified at redshift $z_{\rm id}=0$ and show that the classic 
ellipsoidal collapse model systematically overestimates their collapse times. If one imposes that 
halo collapse takes place at $z_{\rm id}$, this model requires starting from a significantly lower 
linear density contrast than what is measured in the simulations at the locations of halo formation. 
We attempt to explain this discrepancy by testing two key assumptions of the model. First, we show 
that the tides felt by collapsing haloes due to the surrounding large-scale structure evolve 
non-linearly. Although this effect becomes increasingly important for low-mass haloes, accounting 
for it in the ellipsoidal collapse model only marginally improves the agreement with N-body 
simulations. Second, we track the time evolution of the physical volume occupied by forming haloes 
and show that, after turnaround, it generally stabilizes at a well-defined redshift, 
$z_{\rm c}>z_{\rm id}$, contrary to the basic assumption of extended Press-Schechter theory based 
on excursion sets. We discuss the implications of this result for understanding the origin of the 
mass-dependence and scatter in the linear threshold for halo formation. Finally, we show that, 
when tuned for collapse at $z_{\rm c}$, a modified version of the ellipsoidal collapse model that 
also accounts for the triaxial nature of protohaloes predicts their linear density contrast in an 
unbiased way.
\end{abstract}


\section{Introduction}

The formation of dark-matter haloes through gravitational instability
of small density perturbations is a formidable non-linear problem.
Most of our current understanding of the process is based on N-body simulations,
yet valuable theoretical insight can be gained through analytic models that attempt 
to approximate the growth of structure.

Most theoretical models are based on the assumption that haloes of mass $M$ originate from compact 
Lagrangian patches of initial comoving size $R\propto M^{1/3}$ in which conditions are favorable
for collapse. In the most simplistic picture, the collapse process can be approximated by following 
the evolution of a spherically symmetric perturbation with a top-hat density
profile and vanishing initial peculiar velocities
in a otherwise homogeneous and expanding background \citep{Partridge-Peebles-1967, Peebles-1967,
Gunn1972,Peebles1980}. In this ``spherical collapse model'', perturbations that are dense enough
to form bound structures decouple from the background expansion, eventually 
reverse their motion and start collapsing at an increasingly
faster rate. The mathematical solution of the equation of motion
leads to a singularity, but the development of non-radial motions
due to the imperfect symmetry of any realistic perturbation
is expected to halt the collapse and form a stable structure in 
virial equilibrium. Energy conservation suggests that the final radius 
of the bound structure should be approximately one half of the maximum ``turn-around'' radius 
\citep{Gott-Rees-1975}, or smaller in the presence of a cosmological constant \citep{Lahav1991}.

Based on the statistical properties of Gaussian random fields,
\citet{PS1974} developed a model for the number density of dark-matter haloes as a function of their
mass and redshift. The model assumes that all Lagrangian patches in which the
linearly extrapolated density contrast at redshift $z$ lies above
a critical value $\delta_{\rm c}$ (of order unity) have collapsed to form bound haloes 
by that redshift. For practical applications, one has to choose a specific value for
$\delta_{\rm c}$ that matches a given halo definition.
In an Einstein-de Sitter universe, spherical
collapse produces virialized haloes with a final mean overdensity
of $\Delta_{\rm vir}=18\, \pi^2\simeq 178$ which occurs when the
linear density contrast is $\delta_{\rm c}=(3/5)(3\,\pi/2)^{2/3}\simeq 1.686$
\citep{Kaiser84,Bardeen1986,EFWD88}. For different cosmological models,
these quantities acquire a redshift dependence which, however, is very weak
for $\delta_{\rm c}$ \citep[e.g.][]{Eke1996}.

\citet{Bond1991} provided a sounder theoretical basis for the Press-Schechter
mass function using the theory of ``excursion-sets''.
A key assumption of their model is that a halo identified at redshift
$z_{\rm id}$ should form by collecting all matter initially contained
within the largest possible region over which the mean linear overdensity
is $\delta_{\rm L}(z_{\rm id})=\delta_{\rm c}$.
In other words, the Lagrangian boundary of a halo coincides with the
outermost shell which is collapsing at redshift $z_{\rm id}$.
However, for $\delta_{\rm c}\simeq 1.686$,
this method yields halo mass functions that agree only qualitatively
with those extracted from N-body simulations; they exhibit systematic shifts 
from the numerical result at both low and high
masses which can be eliminated by adopting an effective mass-dependent threshold for
halo formation $\delta_{\rm c}(M)$ \citep[e.g.][]{Sheth1999}.

\defcitealias{LudlowPorciani2011b}{Paper I} 	
\defcitealias{Bond1996}{BM96} 	
One possible explanation for this discrepancy is that the spherical
collapse model is too simplistic. For example, N-body simulations have shown that dark-matter
haloes originate from elongated Lagrangian regions whose longest geometric
axis aligns with the direction of maximum gravitational compression
\citetext{\citet{Porciani2002b}, see also \citet{Despali2013,Lee2000,LudlowPorciani2011b}, hereafter Paper I}
The collapse of an ellipsoidal top-hat overdensity amplifies any initial
departure from sphericity whether perturbations are 
isolated \citep{LyndenBell1964,LinMestelShu1965} or embedded in a
uniform and expanding background \citep{Icke1973,WhiteSilk1979,Peebles1980}.
The presence of external tides generated by large-scale
structure, however, is expected to influence the dynamics of collapsing ellipsoids 
\citep[e.g.][]{Hoffman1986,Bertschinger-Jain-1994}.
These tides were incorporated into the EC model by
\citet{EisensteinLoeb1995} and \citet[][hereafter BM96]{Bond1996} in a way that
recovers the Zel'dovich approximation \citep{Zeldovich1970}
in the linear regime. In the latter formulation, an initially spherical overdensity 
is sheared into a collapsing ellipsoid by the action of external tides.
The perturbation first reaches zero extension along the direction of largest 
compression, at which point orbit crossing occurs and the single-stream fluid equations
cease to be valid. This can be prevented, however, by artificially halting 
collapse once the axis has shrunk by a critical
factor $(18 \, \pi^2)^{-1/3}\simeq 0.178$ with respect to the background
expansion \citetext{\citetalias{Bond1996}; \citet{AngrickBartelmann2010}}.

Using this model, and defining the collapse time of a perturbation to be
the epoch at which its last principal axis freezes out,
\citet{Sheth2001} showed that more strongly sheared perturbations
require higher initial density contrasts to overcome the tidal
stretching and collapse by a particular time. Approximating the locations of halo formation as
random points in a Gaussian random field, this ``ellipsoidal-collapse threshold'', $\delta_{\rm ec}$,
can be expressed in terms of the rms amplitude of linear density perturbations, $\sigma(M)$, or 
equivalently in terms of halo mass. The value of $\delta_{\rm ec}$
typically increases towards lower masses in a way that resembles
the measured mean linear overdensities of dark-matter ``protohaloes'' identified in 
the initial conditions of N-body simulations \citetext{\citet{Robertson2009,Elia2012}; \citetalias{LudlowPorciani2011b}}; 
this is generally interpreted as a reflection of the stronger tides felt, on 
average, by less massive haloes.

\citet{Sheth2001} used the ellipsoidal-collapse threshold to predict the
mass function and bias of dark-matter haloes in the excursion-set formalism.
If $\delta_{\rm ec}$ is rescaled by an ad-hoc factor,
this solution offers better agreement with the results of numerical simulations than calculations
based on the spherical collapse model.
\citet{Sheth2001} justified the rescaling by noting that the halo finder that
was used in the simulations did not necessarily match the final overdensity
of the haloes produced by the EC model.

The EC model of \citetalias{Bond1996}, however, cannot account
for the considerable scatter in the measured linear overdensities, $\delta_{\rm L}$, of
regions that later collapse to form haloes of a particular mass $M$,
and additionally fails to explain why $\delta_{\rm L}$ depends strongly on the characteristic
half-mass formation time of the halo \citepalias{LudlowPorciani2011b}.
In that companion paper we showed that, although the average
overdensity of dark-matter protohaloes tends to scale with external
tides as described by the ellipsoidal model of \citetalias{Bond1996},
the majority of recently collapsed haloes fall systematically
{\em below} the model-predicted threshold for collapse.
\citet{Hahn-Paranjape-2014} reached similar conclusions using
warm-dark-matter simulations and thus provided further evidence that the
ellipsoidal model systematically over-predicts the collapse time of a perturbation.

These puzzling results can be explained
by a modified EC model that allows for initial
asymmetry in the shape of the linear perturbations.
Changing the initial axis lengths alters their individual collapse times and
therefore modulates the initial density contrast required for complete
collapse by a particular redshift.
In \citetalias{LudlowPorciani2011b}, we showed that a model tuned to match the Lagrangian
shapes of dark-matter protohaloes in numerical simulations
can accurately reproduce their linear overdensities 
as well as its dependence on the initial departure from sphericity.
A nice feature of this model is that the three principal axes of the
perturbation freeze out almost simultaneously, similar to the case of
spherical collapse. There is one caveat, however: the predicted threshold
for collapse at redshift $z_{\rm id}$ only traces
the overdensities of recently collapsed haloes for a given external tidal
field and lies below the mean value measured
for fixed halo mass and identification redshift.
This begs the question of why dark-matter protohaloes with linear
overdensities substantially above the ellipsoidal-collapse threshold exist at all,
when virtually none with lower initial densities are found in cosmological
simulations. We address these questions here using different versions of the ellipsoidal
collapse model combined with the same numerical simulations (and
dark-matter halo catalogs) as in \citetalias{LudlowPorciani2011b}.
We anticipate that our findings imply a major revision of the standard lore
for halo formation and raise questions regarding the validity of the 
excursion-set ansatz.

This paper is structured as follows. The dynamical model for ellipsoidal
collapse is presented in Section~\ref{ch_model}, while our simulations,
halo catalogs and analysis techniques are discussed in Section~\ref{ch_sim}. 
Our main results are presented in Section~\ref{ch_results}, with a discussion 
of the implications of our results in Section~\ref{ch_discussion}.
The conclusions are then summarized in Section~\ref{ch_summary}.

\section{The Ellipsoidal Collapse Model} \label{ch_model}

The dynamical equations for the collapse of a constant density ellipsoid in the presence
of an external tidal field were derived in \citetalias{LudlowPorciani2011b}. We assume that the principal axes of 
the perturbation are aligned with the eigenvectors of the external tidal field \citep{Porciani2002b}, 
and that the background expansion is driven by a pressureless matter density and a cosmological 
constant (see \cite{DelPopolo2002} for a similar model).

In a Cartesian coordinate system that is aligned with the principal frame of the 
ellipsoid, the differential equation for the axis lengths, $r_i$, is \citepalias{Bond1996}
\begin{eqnarray}
\label{eq_model_differantial}
\frac{\ddot{r}_i}{r_{i}}=\Omega_{\rm \Lambda} \, H^2_0-\frac{3}{2}\, \frac{\Omega_{\rm M} \,
H^2_0}{a^3}\left(\frac{1}{3}+\lambda_i^{\rm tot}\right), 
\end{eqnarray}
where the dots denote time derivatives; $H_0$ is the Hubble constant; $a$ the expansion 
factor; and $\Omega_{\rm M}$ and $\Omega_{\Lambda}$ are the present-day densities of matter and 
the cosmological constant, $\Lambda$, normalized to the critical density, 
$\rho_{\rm crit}=3\,H_0^2/8\,\pi\,G$, where $G$ is Newton's gravitational constant. The tidal 
field, $\lambda_i^{\rm tot}$, is given by 
\begin{eqnarray}
  \label{eq_lambda_total}
  \lambda_i^{\rm tot}=\lambda_i^{\rm ext}+\frac{\delta}{3}+\frac{\beta_{\rm i} \, \delta}{2},
\end{eqnarray}
where $\delta=\delta(a)$ is the time-dependent density contrast of the ellipsoid.
Note that eq.~(\ref{eq_lambda_total}) contains contributions from both external tides, 
$\lambda_i^{\rm ext}$, as well as an internal component generated by the ellipsoid 
itself. The latter can be calculated explicitly using elliptic integrals:
\begin{eqnarray}
  \label{eq_beta}
  \beta_i=r_1\, r_2 \, r_3 \, \int_0^{\infty}\, \frac{{\rm d}\tau}
  {(\tau+r^2_i)\, \prod_{k=1}^3\sqrt{\tau+r^2_k}}-\frac{2}{3} \, , 
\end{eqnarray}
where the $2/3$ guarantees $\sum_i \, \beta_i=0$. Note that for a spherical
geometry, the integral in eq.~(\ref{eq_beta}) is equal to 2/3 and internal tides vanish.
Initial condition for eq.~(\ref{eq_model_differantial}) are set at some early time, $a_0$, 
using the Zel'dovich approximation. 

The external tidal field, $\lambda_i^{\rm ext}$, however, as well as its time evolution, must 
be explicitly specified. One common assumption is that $\lambda_i^{\rm ext}$ is generated
by structure on large scales and evolves from its initial value according to linear theory; 
 another possibility is that $\lambda_i^{\rm ext}$ is dominated by the (non-linear) tidal 
field generated by the perturbation itself \citepalias{Bond1996}. A model which interpolates between
these two regimes was recently proposed by \citet{AngrickBartelmann2010}. In this approach, one
adopts the non-linear model $\lambda_i^{\rm ext}(t)$ until axis $i$ turns around, at which point
its corresponding eigenvalue continues to evolve linearly. In our model we will initially assume 
that external tides are generated by large-scale structure and grow with time according to the 
linear growth factor, $D(z)$; internal tides are calculated self-consistently using the time-dependent
shape of the ellipsoid. In Section \ref{ss_tides} we will revisit the issue of external tides in more 
detail, in order to test the assumption made above regarding their time evolution.

The collapse and virialization of the perturbation is generally approximated in this model by freezing 
the individual axes when they reach a size $r_{f,i}=f\, q_i\, a$ \citepalias{Bond1996}. Here $q_i$ is the {\em initial} comoving
length of axis $i$, and the parameter $f$ is usually set to 0.178. In the case of spherical collapse
in an Einstein-de Sitter universe, this choice ensures that the perturbation has a density contrast 
of $\delta\approx 178$ at the moment of collapse. Note, however, that a particular choice of $f$ has
no fundamental physical motivation and may depend on the nature of collapse (e.g., whether spherical
or ellipsoidal) or on the background cosmological model. A more general virialization condition 
based on the tensor virial theorem was suggested by \citet{AngrickBartelmann2010}. However, since the 
late stages of collapse generally occur quite rapidly, such modifications have only a minor effect on 
axis collapse times, from which density thresholds are inferred. For the sake of simplicity, and to ease 
comparison with previous work, we adopt the traditional freezing factor $f=0.178$ for our EC model as well, 
but return this point in Section~\ref{ss_zcoll}.

\section{Numerical Methods} \label{ch_sim}

\subsection{The Simulations} \label{ss_sim}

Our analysis focuses on dark-matter haloes identified at $z_{\rm id}=0$ in two cosmological 
simulations of structure formation in the standard $\Lambda$CDM cosmology. These simulations 
are the same as those described in \citetalias{LudlowPorciani2011b} \citep[see also][]{Pillepich2010}. We therefore 
summarize here only their main aspects, and refer the reader to that work for further details. 

Both runs followed the evolution of the dark-matter component using 1024$^3$ equal mass particles. 
The periodic boxes have side lengths equal to 150 $h^{-1}$Mpc and 1200 $h^{-1}$Mpc. Each run was 
carried out with a lean version of the simulation code {\sc gadget} \citep{Springel2001a} and adopted 
the following cosmological parameters: 
$\Omega_M=0.279$, $\Omega_{\Lambda}=1-\Omega_{M}=0.721$, $n_s=0.96$, $\sigma_8=0.817$, and 
$H_0=73$ km s$^{-1}$ Mpc$^{-1}$. Here $n_s$ is the spectral index of primordial density 
fluctuations; $\sigma_8$ is the rms density fluctuation measured in 8 $h^{-1}$ Mpc spheres, linearly 
extrapolated to $z=0$. These values are consistent with the WMAP 5-year data release 
\citep{Komatsu2009}. The resulting particle masses are $m_p=2.43\times10^8\, h^{-1}$ M$_\odot$ and 
$m_p=1.24\times10^{11} \, h^{-1}$ M$_\odot$ for the 150 $h^{-1}$Mpc and 1200 $h^{-1}$Mpc boxes.

Initial conditions for the simulations where produced using the Zel'Dovich 
  approximation for a starting redshift of $z_{\rm start}=70$ and $50$ for the small and large 
  box, respectively. As discussed in detail by \citet{Pillepich2010}, these choices of 
  $z_{\rm start}$ are sufficient to erase all transient artifacts in the halo mass function 
  by $z=0$. During each simulation, 30 snapshots were saved between $z=10$ and 0 in logarithmically 
  spaced intervals of $(1+z)^{-1}$.

\subsection{Halo Catalogs} \label{ss_haloes}

We identified dark-matter haloes in each simulation output using a friends-of-friends (FOF) halo 
finder with a linking length equal to 0.2 times the mean nearest-neighbour distance. All haloes with at 
least N$_{\rm FOF}=32$ particles were included in the halo catalogs. Once all dark-matter haloes were
identified, the formation histories of $z_{\rm id}=0$ haloes were constructed by tracing each halo's most 
massive progenitor backwards through all previous simulation outputs. Accretion histories defined 
in this way can be used to provide simple estimates of halo formation times, such as the redshift 
$z_{50}$ at which 50 per cent of the halo's final mass had first assembled into one main progenitor. 

We will also consider the Lagrangian ``protohaloes'' of each $z=0$ halo, which can be easily 
identified by tracing all halo particles back to the initial conditions of the simulation.
In this paper we will only consider haloes that, at $z=0$, are composed of at least N$_{\rm FOF}=1000$ 
particles, unless explicitly stated otherwise.

In order to test the sensitivity of our results to the adopted halo definition, we have 
also built halo catalogs using a spherical overdensity (SO) halo finder. This algorithm grows 
spheres around local density maxima until they reach a density contrast of $\Delta$
times the mean matter density. We adopted $\Delta=100$, 200 and 500, and repeated all aspects of 
the analysis using these alternative halo definitions. We will distinguish the characteristics of 
haloes identified by the FOF or SO algorithms using subscripts. The FOF halo mass, for 
example, will be denoted $M_{\rm FOF}$, whereas $M_{200}$ defines the SO mass based on an overdensity 
of $\Delta=200$. The main results of our work will be presented for the FOF halo definition; the results 
for SO haloes are summarized in Appendix \ref{app_so}.

\subsection{Analysis Techniques} \label{ch_analysis}

The EC model described in Section~\ref{ch_model} depends explicitly on the initial
axis ratios of the collapsing overdensity. Since in practice these are free parameters, we will 
constrain their values using the shapes of protohaloes identified in the initial conditions of our
simulation.

We characterize the shapes of dark-matter haloes and protohaloes using the inertia tensor, 
defined
\begin{eqnarray}
  \label{eq_inertia}
  I_{ij}=m_p \sum_k \, x_{k,i} \, x_{k,j} \, ,
\end{eqnarray}
where ${\bf x}_k$ is the distance vector between particle $k$ and the halo's center-of-mass; the $i$ 
and $j$ are the projected lengths of ${\bf x}_k$ along each coordinate direction. This matrix can be 
diagonalized to obtain the principal axis lengths of the ellipsoid, $q_1\geq q_2\geq q_3$, which can 
be used to characterize halo shapes in terms of their ratios: $q_2/q_1$ and $q_3/q_1$, for example, 
are the intermediate-to-major and minor-to-major axis ratios. The eigenvectors of the inertia tensor 
define the principal axis frame, and will be denoted $\mathbfit{i}_i$.

As discussed in \citetalias{LudlowPorciani2011b}, the shapes of protohalo boundaries are closely 
related to the external tidal field acting upon them. Protohaloes have strongly triaxial shapes whose
principal directions align closely with those of the surrounding tides. When estimating the tides
acting upon a given protohalo, it is therefore desirable to move beyond simple spherical filtering 
of the tidal field, and attempt to incorporate additional information on the shape of the protohalo
as well.

Here we propose a novel method to estimate the {\em average} tidal field acting upon a protohalo as it 
collapses to form a non-linear object. We start by calculating the density contrast field on a 
1024$^3$ grid that covers the entire simulation volume. Densities are assigned to each grid element 
using cloud-in-cell interpolation \citep{HockneyEastwood1988}. Within each grid element, we also 
calculate the tidal deformation tensor,
\begin{eqnarray}
  {{D}}_{ij}=\frac{\partial^2\Phi}{\partial x_i \partial x_j},
  \label{deform}
\end{eqnarray}
using standard Fourier techniques. Here $\Phi$ is the peculiar gravitational potential, 
which is related to the density contrast by Poisson's equation: $\nabla^2\Phi=\delta$.
This field can be used compute the total strength and orientation of the tidal field over an arbitrary volume, 
$V$ (for example, a protohalo), in the following way. We first compute the magnitude of $D_{ij}$ along the direction 
$\boldsymbol{\eta}$ using
\begin{eqnarray}
  \label{eq_tidal_average}
  \lambda=\sum_V \boldsymbol{\eta}\cdot \mathbfss{D} \cdot\boldsymbol{\eta} \, ,
\end{eqnarray}
where the sum is over all grid cells contained within the volume. We then iteratively determine 
the orientation of $\boldsymbol{\eta}$ that minimizes $\lambda$; this defines the direction 
$\mathbfit{d}_3$ and its magnitude $\lambda_3$. In the plane perpendicular to $\mathbfit{d}_3$, 
we then search for the direction $\boldsymbol{\eta}=\mathbfit{d}_1$ that maximizes 
eq.~(\ref{eq_tidal_average}), which also determines $\lambda_1$. This constrains the direction 
$\mathbfit{d}_2$ since it is, by definition, perpendicular to both $\mathbfit{d}_1$ and $\mathbfit{d}_3$; 
$\lambda_2$ is then determined using eq.~(\ref{eq_tidal_average}) along the direction 
$\boldsymbol{\eta}=\mathbfit{d}_2$. In this way we estimate the total tidal tensor, described by 
the $\lambda_i$s and $d_i$s, acting on the volume $V$ without invoking a spherical filter function.
For protohaloes, we assume that $V$ can be approximated by an ellipsoid whose axis lengths 
and orientations are derived from its inertia tensor.

Once the linear tidal field has been measured in this way, we define the Lagrangian overdensity
of each protohalo as $\delta_{\rm L}=\sum_i \, \lambda_i$. Both the tides and the density contrast 
of the protohaloes are therefore evaluated within an ellipsoidal (rather than spherical) aperture 
whose shape and orientation is tailored to match each individual protohalo. Relative to spherical 
filtering, this results in per cent-level corrections to $\delta_{\rm L}$ over the range of masses 
we study here.

Applying this procedure to FOF haloes with $\geq$1000 particles ensures that even the lowest mass 
haloes are, on average, resolved with at least 20 grid cells. We have explicitly verified that this 
is sufficient to yields robust estimates of the $\lambda_i$s and $\mathbfit{d}_i$s.

Note that the tidal field can be alternatively characterized in terms of its ellipticity, $e$, and 
prolateness, $p$. These are defined
\begin{eqnarray}                                                      
e\ = \frac{\lambda_1-\lambda_3}{2\, \delta_{\rm L}}
\label{eq:ell}                                                              
\end{eqnarray}
and
\begin{eqnarray}                                                                                                      
p\ = \frac{\lambda_1-2\, \lambda_2+\lambda_3}{2\, \delta_{\rm L}}. 
\label{eq:prl}                                                                                                        
\end{eqnarray}

\section{Results} \label{ch_results}

\subsection{Ellipsoidal collapse and density thresholds for CDM halo formation} \label{ss_simellip}

%%%%%%%%%%%%%%%%%%%%%%%%%%%%%%%%%%%%%%%%%%%%%%%%%%%%%%%%%%%%%%%%%%%%%%%%%%
\begin{figure*}
  \begin{center}
    \resizebox{16cm}{!}{\includegraphics{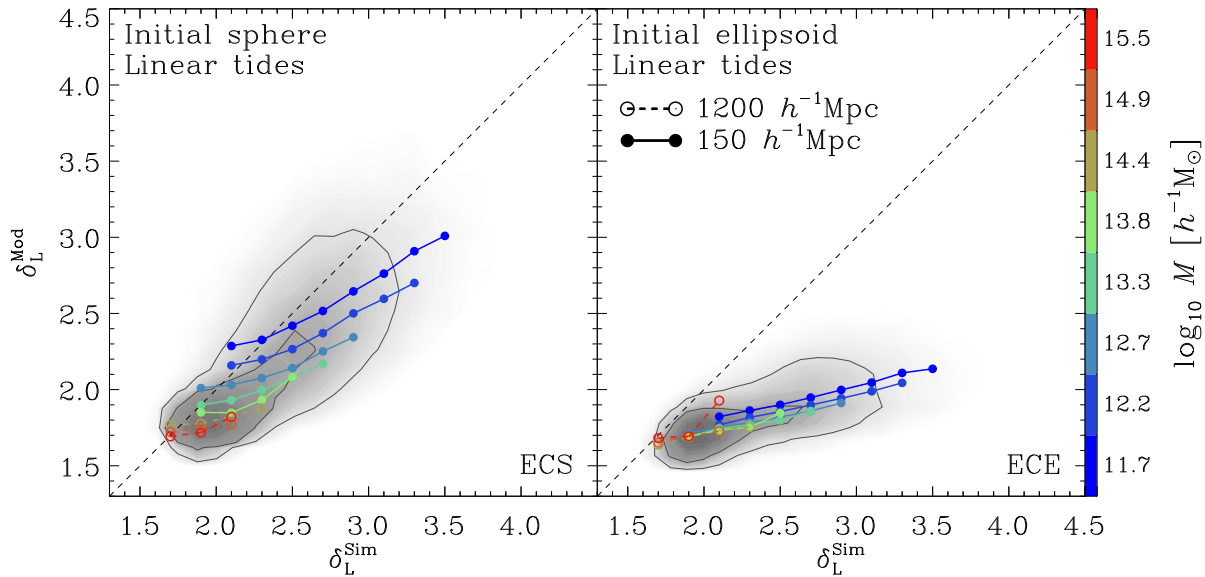}}
  \end{center}
  \caption{Lagrangian overdensities of dark-matter protohaloes predicted by the ellipsoidal 
    collapse model plotted versus their overdensities measured in the initial conditions
    of our simulations. Model predictions assume that the Lagrangian tidal field measured
    at each protohalo's center evolves according to linear theory. The left-hand panel
    corresponds to the EC model of \citetalias[][ECS]{Bond1996}, which assumes 
    that each protohalo's initial shape can be approximated by a sphere; the right-hand 
    panel explicitly accounts for each protohalo's initially non-spherical shape, which we 
    measure in the simulation initial conditions. Shaded regions highlight the density of points 
    in the $\delta-\delta$ plane; contours enclose 50 per cent and 80 per cent of the data. 
    Connected points show the median values of $\delta_{\rm L}^{\rm Mod}$ at fixed 
    $\delta_{\rm L}^{\rm Sim}$ measured in separate mass bins (shown using different colored lines), 
    equally spaced in $\log M$. Solid circles are used for haloes in our 150 $h^{-1}\,$ Mpc box; 
    open squares for those in the 1200 $h^{-1}\,$ Mpc box}
  \label{pic_dd_all_today}
\end{figure*}
%%%%%%%%%%%%%%%%%%%%%%%%%%%%%%%%%%%%%%%%%%%%%%%%%%%%%%%%%%%%%%%%%%%%%%%%%% 

Given estimates of the tidal field acting upon a given protohalo, we can use the 
EC model described in Section~\ref{ch_model} to {\em predict} the linear 
density contrast required for collapse to occur at $z_{\rm id}=0$. We will discriminate
``predicted'' and ``measured'' values of $\delta_{\rm L}$ using superscripts: 
$\delta_{\rm L}^{\rm Sim}$, for example, refers to its value measured in the initial
conditions of our simulations; $\delta_{\rm L}^{\rm Mod}$ to the model-predicted value.
Protohalo overdensities predicted by the EC model are shown in Figure~\ref{pic_dd_all_today}, 
and are compared directly with the linear overdensities $\delta_{\rm L}^{\rm Sim}$ obtained by 
the method described above. We will use these ``$\delta-\delta"$ relations as a diagnostic for the 
ability of the EC model to describe the dynamics of individual dark-matter haloes.

In Figure~\ref{pic_dd_all_today} we plot, for each protohalo, the linear density contrast
for collapse at $z=0$ predicted by the EC model versus their measured 
overdensities. The left hand panel shows the predictions of the EC model 
of \citetalias{Bond1996}, which assumes that each protohalo occupies a spherical Lagrangian volume\footnote{
Note that an equivalent plot was provided in Figure 3 of \citet{Sheth2001}.}.
We will hereafter refer to this as the ECS model (for Ellipsoidal Collapse of Spherical 
perturbations); density contrasts predicted by this model will be denoted $\delta_{\rm L}^{\rm ECS}$. 
On the right, we have included the influence of each protohalo's shape on the predicted 
collapse threshold (hereafter referred to as the ECE model, for Ellipsoidal Collapse of Ellipsoidal 
perturbations). Note that, in this case, the non-spherical perturbation itself
contributes to the initial tidal field. We therefore modify the external component
such that the total tidal field in the model initially matches the linear tidal field 
measured for each protohalo in the simulation. In each panel connected symbols show the 
median values of $\delta_{\rm L}^{\rm Mod}$ measured in fixed bins of $\delta_{\rm L}^{\rm Sim}$; 
different colored lines plot the relations for equally spaced logarithmic mass bins (indicated in 
the legend).

%%%%%%%%%%%%%%%%%%%%%%%%%%%%%%%%%%%%%%%%%%%%%%%%%%%%%%%%%%%%%%%%%%%%%%%%%% 
\begin{figure*}
  \begin{center}
    \resizebox{0.7\textwidth}{!}{\includegraphics{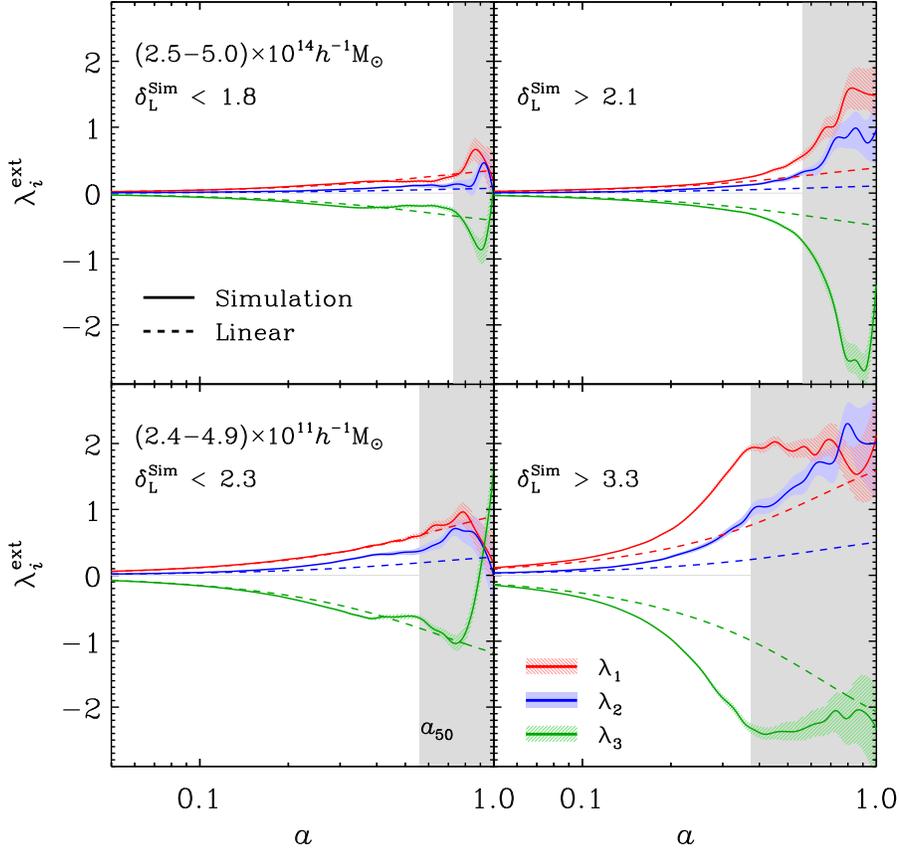}}
  \end{center}
  \caption{Evolution of the median external tidal field (solid lines) for haloes in two separate
    mass bins. Upper panels correspond to haloes in the mass range $2.5<M_{\rm FOF}/(10^{14} \, h^{-1} \, M_\odot) < 5$,
    lower panels to those with $2.4<M_{\rm FOF}/(10^{11} \, h^{-1} \, M_\odot) < 4.9$.
    Left- and right-hand panels split haloes in each mass bin according to their initial
    density contrast, $\delta_{\rm L}^{\rm Sim}$: those of the left include only haloes that rank in the lowest 
    15 per cent of the $\delta_{\rm L}^{\rm Sim}$ distribution, and those on the right only those in the highest 
    15 per cent. In each case, hatched regions correspond to the 90 per cent confidence interval 
    on the median $\lambda_i^{\rm ext}(z)$ obtained by bootstrapping. For comparison, we also show 
    the linear evolution of the median Lagrangian tidal fields measured at each halo center
    using dashed lines. Grey shaded regions correspond to redshifts $z<z_{50}$, where 
    $z_{50}$ is the median half-mass formation redshift of each halo sample.}
  \label{pic_tides_nonlinear_ratio}
\end{figure*}
%%%%%%%%%%%%%%%%%%%%%%%%%%%%%%%%%%%%%%%%%%%%%%%%%%%%%%%%%%%%%%%%%%%%%%%%%% 

Note that, in both cases, the model-predicted overdensities correlate rather well with those 
measured directly in the initial conditions of the simulations, albeit with considerable scatter. 
The median trends, however, are noticeably shallower than one would expect if the ellipsoidal 
model truly captures the dynamics of halo collapse. Note also that the ECS model predicts a 
strong mass dependence to the median $\delta-\delta$ relations. This results from the fact that 
the collapse barrier (at fixed $p$) depends entirely on the ellipticity of the tidal field $e$, 
growing monotonically with increasing $e$. Because, for random points, $e$ scales with mass 
as $e=(\sigma(M)/\delta)/\sqrt{5}$  \citep{Doroshkevich1970}, the ECS model predicts systematically 
higher collapse thresholds for lower mass haloes, resulting in a segregation of the average $\delta_{\rm L}^{\rm Mod}$s 
predicted for haloes of different mass.

Intriguingly, the mass dependence disappears when individual protohalo shapes are 
included in the model prediction. This is because toward lower masses, protohaloes become 
increasingly triaxial, which lowers the density threshold required for collapse
to occur in spite of the increasing tidal field strength \citepalias[see][]{LudlowPorciani2011b}. None the less, both models 
fail to reproduce the measured distribution of protohalo overdensities, and it is worthwhile 
exploring what aspects or assumptions of the EC model may result in the 
discrepancy.

\subsection{The influence of external tides} \label{ss_tides}

%%%%%%%%%%%%%%%%%%%%%%%%%%%%%%%%%%%%%%%%%%%%%%%%%%%%%%%%%%%%%%%%%%%%%%%%%%
\begin{figure}
  \begin{center}
    \resizebox{8cm}{!}{\includegraphics{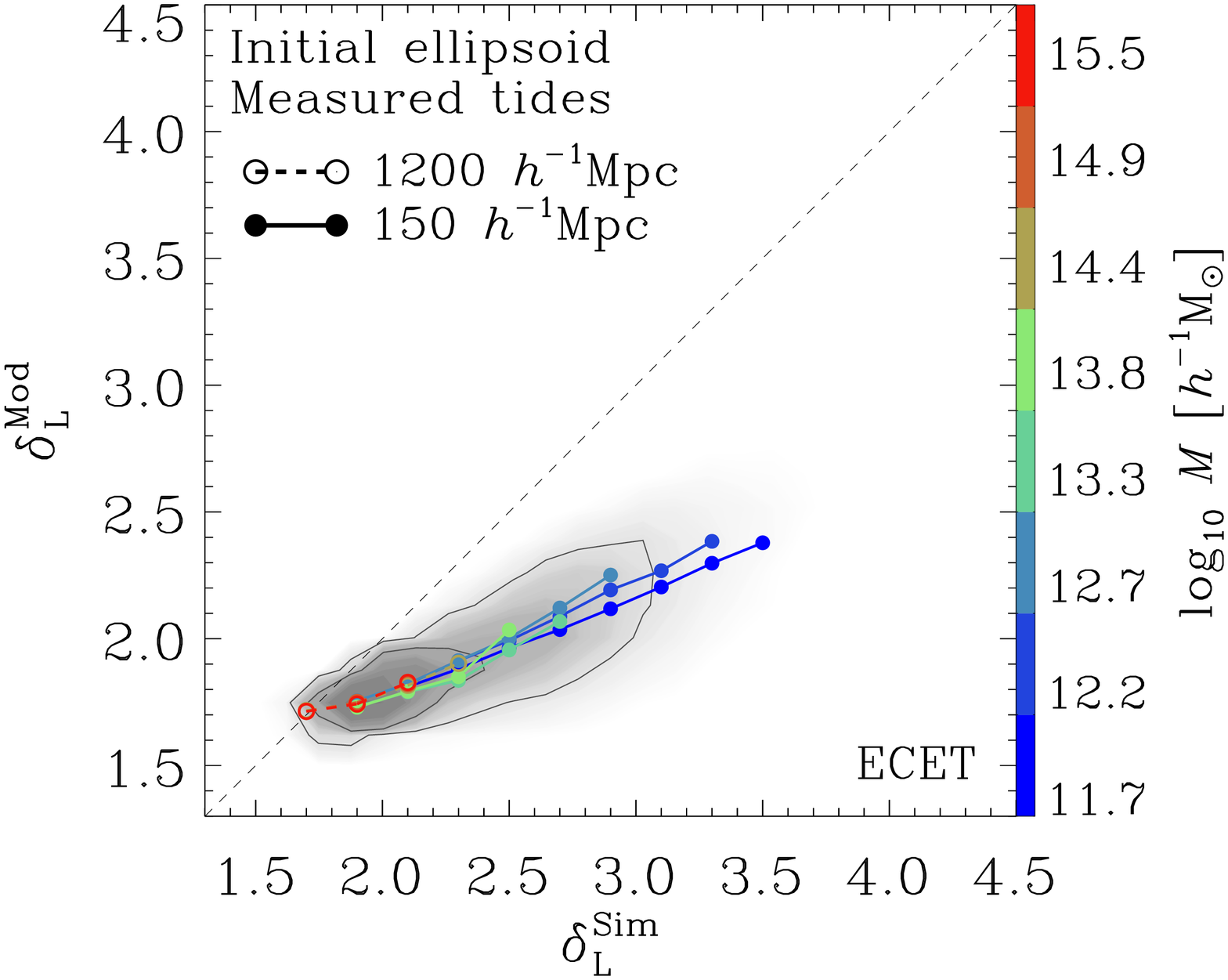}}
  \end{center}
  \caption{Linear density contrast predicted by the ECET model plotted against the 
    measured Lagrangian overdensities of dark-matter protohaloes. The ECET model explicitly
    accounts for the non-spherical shape of each individual protohalo as well as 
    the evolution of their external tidal fields, without resorting to 
    the common assumption of linearly evolving tides. As in Figure~\ref{pic_dd_all_today},
    shaded regions indicate the density of haloes in the $\delta-\delta$ plane;
    contours enclose 50 per cent and 80 per cent of the data points. Connected points show
    the median values of $\delta_{\rm L}^{\rm Sim}$ at fixed $\delta_{\rm L}^{\rm Mod}$ 
    measured in separate mass bins, as indicated in the legend.}
  \label{pic_dd_all_today_ECET}
\end{figure}
%%%%%%%%%%%%%%%%%%%%%%%%%%%%%%%%%%%%%%%%%%%%%%%%%%%%%%%%%%%%%%%%%%%%%%%%%% 

One common assumption of the EC model -- and indeed our assumption in 
constructing Figure~\ref{pic_dd_all_today} -- is that the external tidal field,
assumed to be generated by structure on very large scales, remains linear at all times. 
All ingredients needed to solve eq.~(\ref{eq_model_differantial}) are therefore already 
present in the linear density field. In reality, a growing dark-matter halo may be 
subjected to interactions with nearby neighbours which may result in strongly non-linear 
tidal forces that act to suppress its growth and alter its collapse time
\citep{Hahn2009,WangMoetal2011}. It is therefore 
important to assess whether the assumption of linearly evolving tides remains valid during the 
evolution of simulated haloes, in order to make a more meaningful comparison between their 
measured and predicted overdensities.

To do so, we trace all particles belonging to each protohalo through each simulation output, 
and use eq.~(\ref{eq_inertia}) to characterize the redshift dependence of the shape and orientation 
of each collapsing region. We approximate the geometry of this region as an ellipsoid and fix its
volume, $V$, such that it encloses the protohalo mass at all subsequent times. The total tidal field
acting upon the collapsing region is then estimated using the procedure outlined in Section 
~\ref{ch_analysis}. This method has two distinct advantages over using the particles themselves 
to define $V$: 1) it ensures that we follow a region with constant enclosed mass, even
though the individual particles within it may change with time; 2) it allows for a simple
decomposition of the total tidal field into its internal and external components.

We apply this 
procedure to each $z_{\rm id}=0$ FOF halo (containing at least 1000 particles) and in all simulations 
outputs in order to explicitly measure the time evolution of the tidal field, $\lambda_i^{\rm tot}(z)$, 
acting upon each halo. The shape of the ellipsoid can be used to approximate the contribution of 
internal tides, $\beta_i\delta/2$, using eq.~(\ref{eq_lambda_total}) and (\ref{eq_beta}), in order to 
estimate $\lambda_i^{\rm ext}(z)$. Note that calculating the internal tides this way assumes a homogeneous 
density inside the ellipsoid, which may be inaccurate at late times.

The evolution of the external tidal field calculated in this way is shown in Figure 
\ref{pic_tides_nonlinear_ratio}  for haloes in two separate mass bins. Upper panels correspond
to haloes that fall in the mass range $2.5<M_{\rm FOF}/(10^{14} \, h^{-1} \, M_\odot) < 5$; 
the lower to those with $2.4<M_{\rm FOF}/(10^{11} \, h^{-1} \, M_\odot) < 4.9$. Panels on the 
left show haloes in each mass bin that rank in the lowest 15 per cent of $\delta_{\rm L}^{\rm Sim}$, 
whereas those on the right rank in the highest 15 per cent. (The threshold values of 
$\delta_{\rm L}^{\rm Sim}$ for each sample are provided in each panel.) 
Grey shaded regions indicate redshifts below the median half-mass formation redshift for each sample, after 
which the assumption of a homogeneous density inside the ellipsoid is likely inaccurate.
The measured $\lambda^{\rm ext}_i(z)$s are shown using solid lines, as indicated in the legend, 
while the dashed lines show the linear extrapolation of the average Lagrangian tides measured 
for each halo sample.

Massive haloes tend to evolve in environments where the tidal field scales approximately
as expected from linear theory. This is not surprising given that these haloes 
dominate their environments and the external tides are therefore generated by long-range modes
that have not yet collapsed. The same is true for low-mass haloes in low-density
regions. Protohaloes with low values of $\delta_{\rm L}^{\rm Sim}$ tend to to be less clustered 
than those of higher $\delta_{\rm L}^{\rm Sim}$, with the latter forming in regions 
where tidal fields clearly grow more strongly than expected from a simple
linear extrapolation of the Lagrangian values. 

What implications do these non-linear tides have for the collapse thresholds of dark-matter haloes 
inferred from the EC model? Having calculated the evolution of the tidal field explicitly for each 
individual halo, we can insert it directly into eq.~(\ref{eq_model_differantial}) in order to asses 
the impact of non-linear tides on the collapse thresholds inferred from the ellipsoidal collapse 
model. The resulting $\delta-\delta$ relation for the ECE model is show in Figure~\ref{pic_dd_all_today_ECET}
(hereafter the ECET model) where, again, the connected points highlight medians in logarithmic
mass bins. 

As expected, massive haloes, and those with $\delta_{\rm L}^{\rm Sim}\approx\delta_{\rm sc}$, are largely 
unaffected by the inclusion of explicitly measured external tides. Those with higher 
$\delta_{\rm L}^{\rm Sim}$, however, live in more clustered environments and are affected by tides 
that clearly evolve non-linearly. These tides act to inhibit the collapse of the 
density perturbations, increasing the initial density contrast required for collapse 
to occur at $z=0$. The effect, however, is weak. For example, haloes in the lower 
right panel of Figure~\ref{pic_tides_nonlinear_ratio} are subject to external tides 
that deviate from the linear theory extrapolation by roughly a factor of $\sim$2.2 at the 
halo's half-mass formation time (shaded regions indicate redshifts below the median 
value of $z_{50}$ for all haloes in each sample). This effect 
delays full collapse (in the ECE case) by roughly $1.4\, h^{-1}$Myr, or, equivalently, 
requires an enhancement of only 12 per cent in the initial density contrast for collapse 
at $z=0$. This is clearly not sufficient to bring the predictions of the ECE model 
into agreement with the simulation data, suggesting that other factors may be at play. 
We turn our attention to these next.

\subsection{Collapse times} \label{ss_zcoll}

There is a clear mismatch between the predictions of the EC model and the true properties 
of dark-matter protohaloes when applied on a object-by-object basis. As discussed in \citetalias{LudlowPorciani2011b}, 
these differences may be related to the details of each halo's unique evolutionary history. 
For example, in the EC or SC model, collapse occurring at redshift $z_{\rm c}>z_{\rm id}$ requires 
an initial density contrast a factor of $D(z_{\rm id})/D(z_{\rm c})$ larger than what would be needed 
for collapse at $z_{\rm id}$ under the same environmental conditions. Because, at a given mass scale, 
haloes with higher $\delta_{\rm L}$ typically form earlier than those of lower $\delta_{\rm L}$ 
\citepalias{LudlowPorciani2011b}, correcting the model predictions for $z_{\rm c}>0$ may explain the 
discrepancy between the measured overdensities of protohaloes and the model-predicted values.

As discussed in \citetalias{LudlowPorciani2011b}, simple estimates of halo formation times based on the
growth of their most massive progenitors fail to account for the scatter in their linear overdensities. 
However, unlike our simulated haloes, which grow through a sequence of mergers and smooth accretion, 
perturbations in the EC model remain homogeneous throughout their evolution and the 
collapse redshift is therefore unambiguously defined as the time at which the last axis reaches the radius 
$r_{f,i}=f\, q_i\, a$. Once collapse has occurred, the model assumes that the volume of the ellipsoid
remains constant thereafter. Is there an analogous definition of ``collapse'' that can be easily 
applied to cosmological haloes?

For a dark-matter halo identified at $z_{\rm id}=0$, one can approximate the evolution of its 
outermost mass shell by tracking the volume $V(a)$ of the same best-fitting ellipsoid used to 
calculate the external tidal field (described in Section~\ref{ss_tides}). Its size can be used 
to provide a simple and intuitive estimate of the time at which its {\em entire} $z=0$ mass had 
first assembled into a single, dispersion-supported non-linear system.

%%%%%%%%%%%%%%%%%%%%%%%%%%%%%%%%%%%%%%%%%%%%%%%%%%%%%%%%%%%%%%%%%%%%%%%%%% 
\begin{figure}
  \begin{center}
    \resizebox{8cm}{!}{\includegraphics{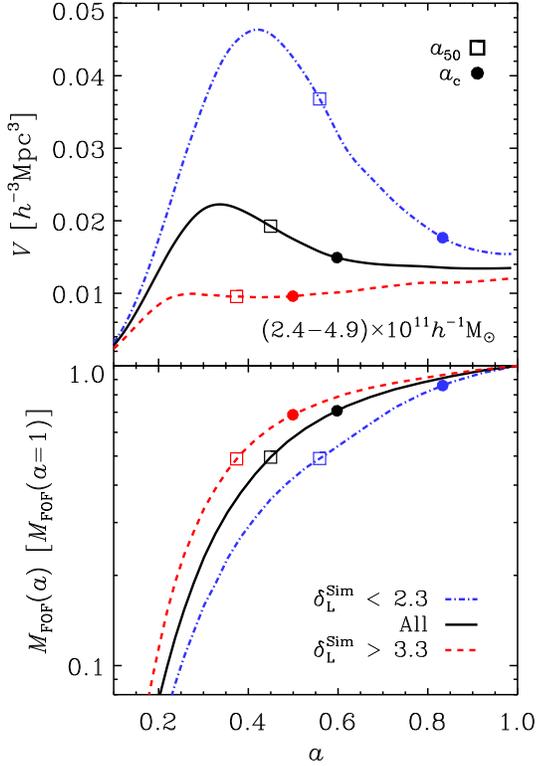}}
  \end{center}
  \caption{ {\em{Upper panel:}} Evolution of the median volume of the ellipsoid, centered on each 
    collapsing protohalo, that, at any $z$, encloses the final mass of the descendant halo at $z=0$
    (see Section~\ref{ss_zcoll} for details). All haloes are selected to lie in a narrow mass bin, 
    as indicated in the legend. The black solid line corresponds to the median $V(a)$ for all haloes 
    in that mass bin. Dashed (red) and dot-dashed (blue) lines show the evolution for the subsamples
    of haloes that lie in the upper-most and lower-most 15 per cent of $\delta_{\rm L}^{\rm Sim}$. 
    {\em{Lower panel:}} Evolution of the median FOF mass of the most massive
    progenitor for the same samples of haloes. In both panels, solid circles highlight the ``collapse
    redshift'', $z_{\rm c}$, defined in Section~\ref{ss_zcoll}, whereas open squares indicate the 
    half-mass formation time, $z_{50}$, at which each haloes main progenitor first assembled half
    of its present-day mass.}
  \label{pic_coll_acc}
\end{figure}
%%%%%%%%%%%%%%%%%%%%%%%%%%%%%%%%%%%%%%%%%%%%%%%%%%%%%%%%%%%%%%%%%%%%%%%%%% 

In the upper panel of Figure \ref{pic_coll_acc} we plot the evolution of the volume of 
three such ellipsoids after averaging over subsamples of haloes in the mass range 
$(2.4-2.9)\times 10^{11}\, h^{-1} \, M_{\odot}$. The solid (black) curve shows the 
evolution of the median $V(a)$ computed for all haloes in the quoted mass bin; dashed (red) and 
dot-dashed (blue) lines show the corresponding result for haloes that rank in the 
maximum and minimum 15 per cent of $\delta_{\rm L}^{\rm Sim}$, respectively.

On average, the present-day mass of these haloes was already in place at $z\approx 0.7$.
More extreme examples can be found in the sample with the highest initial density contrasts,
whose present-day masses reached a stable volume at $z\approx 2$, when the Universe was only 
$\sim$15 per cent of its current age. Attempting to predict the collapse dynamics of these haloes 
with models tuned for collapse at $z=0$ is therefore prone to systematic errors. On the other 
hand, haloes with the lowest value of $\delta_{\rm L}^{\rm Sim}$ appear to have accreted their 
outer-most mass shells only very recently. This implies that, on average, haloes with 
$\delta_{\rm L}^{\rm Sim}>\delta_{\rm L}^{\rm ECE}$ have $z_{\rm c}>0$: strictly speaking, 
{\em these haloes collapsed before they were identified}.

This may seem to conflict with the mass accretion histories of the same haloes, shown in the bottom panel of Figure~
\ref{pic_coll_acc}. These curves trace the median evolution of
the FOF mass of each halo's main progenitor, and suggest that halo masses increase at all redshifts.
However, as already noted by \citet{Diemand2007}, this ``pseudo-growth'' results from the fact that halo
boundaries are defined, at any time, relative to a fixed (or slowly varying) overdensity threshold. 
The decrease in the cosmic background density with time therefore results in an artificial increase of halo 
boundaries, and hence masses \citep[see also,][]{Cuesta2008,Diemer2013,Zemp2013}. Halo finders based on fixed 
{\em physical} densities may therefore result in more realistic estimates of their masses and sizes.

None the less, we can use the trajectories of $V(a)$ to estimate an appropriate ``freezing'' or 
collapse time, $z_{\rm c}$, for each individual halo. This estimate of $z_{\rm c}$ can then be used 
to halt collapse in the EC model in order to make a more appropriate comparison between the 
model's predictions and measured Lagrangian overdensities of dark-matter protohaloes. 

%%%%%%%%%%%%%%%%%%%%%%%%%%%%%%%%%%%%%%%%%%%%%%%%%%%%%%%%%%%%%%%%%%%%%%%%%% 
\begin{figure}
  \begin{center}
    \resizebox{8cm}{!}{\includegraphics{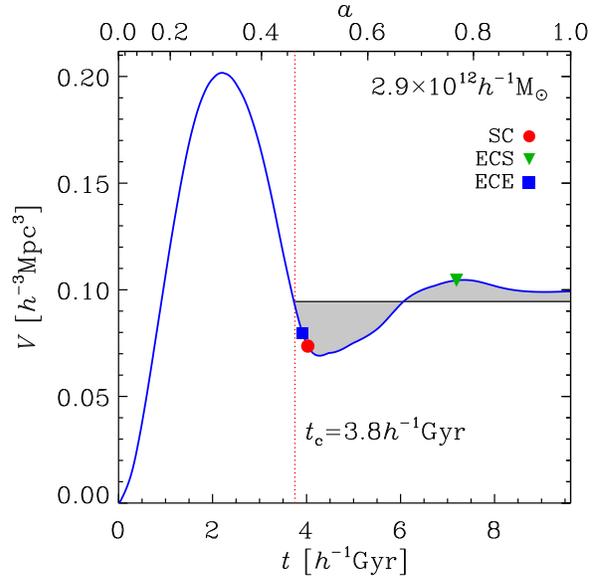}}
  \end{center}
  \caption{Evolution of the volume of an ellipsoid centered on the progenitor of a 
    ($z=0$) $2.8\times 10^{12} \, h^{-1}\, M_\odot$ halo that, at each redshift, encloses the 
    present day halo mass. The shape and orientation 
    of the best-fitting ellipsoid is explicitly calculated using the full set of particles that, at $z=0$,
    belong to the FOF group. The evolution of the volume, $V(z)$, is used to estimate the 
    ``collapse redshift'' of the halo, $z_{\rm c}=1.2$ (dotted-dashed vertical line), using the condition 
    specified in eq.~(\ref{eq_collapse_time}). This definition of $z_{\rm c}$ estimates the time at which
    the entire $z=0$ halo mass had first assembled into a stable configuration, and accounts for
    oscillations in the volume that occur during the system's approach toward equilibrium.
    The colored symbols plotted along the blue line highlight the collapse times  
      predicted by the different versions of the EC model discussed in this paper.}
  \label{ex_freez_mass}
\end{figure}
%%%%%%%%%%%%%%%%%%%%%%%%%%%%%%%%%%%%%%%%%%%%%%%%%%%%%%%%%%%%%%%%%%%%%%%%%% 

We define $z_{\rm c}$ as earliest redshift at which the following condition is satisfied: 
\begin{eqnarray}
  \label{eq_collapse_time}
  \int_{t_{\rm c}}^{t_{\rm max}} [V(t) - V(t_{\rm c})]\, {\rm d}t = 0\, ,
\end{eqnarray}
where $t_{\rm c}$ and $t_{\rm max}$ are the cosmological times corresponding to the collapse redshift, 
$z_{\rm c}$, and the end of the simulation, $z_{\rm max}$. In Figure~\ref{ex_freez_mass} 
we provide an example of the evolution of $V(a)$ (solid blue curve) for a single 
halo of ($z=0$) mass $M_{\rm FOF}\sim 2.9\times 10^{12} \, h^{-1}\, M_\odot$.
The evolution mimics the expectations of simple collapse models: after an initial phase of expansion 
the system turns around, begins collapsing and eventually reaches a stable quasi-equilibrium 
configuration. The first ``dip'' after turnaround corresponds to a state of maximal compression 
in which the halo's entire $z=0$ mass is briefly confined within a compact volume smaller than 
the present-day virial volume of the halo. This is followed by a phase of expansion, and a second 
(very slight) phase of contraction as the system moves toward equilibrium. These ``dips'' and 
``peaks'' clearly occur {\em after} collapse, but before virialization. Our definition of 
$z_{\rm c}$ allows for these oscillations as the system relaxes to a state of equilibrium. The 
grey shaded region in Figure~\ref{ex_freez_mass} highlights the integrand of eq.~(\ref{eq_collapse_time}). 
Note that areas above and below the horizontal line cancel so that the integral vanishes. The 
corresponding collapse time, $z_{\rm c}\approx 1.2$, is indicated with a vertical (red) dotted 
line.

%%%%%%%%%%%%%%%%%%%%%%%%%%%%%%%%%%%%%%%%%%%%%%%%%%%%%%%%%%%%%%%%%%%%%%%%%% 
\begin{figure}
  \centering
  \includegraphics[width=0.48\textwidth,bb=0 0 604 566,keepaspectratio=true]{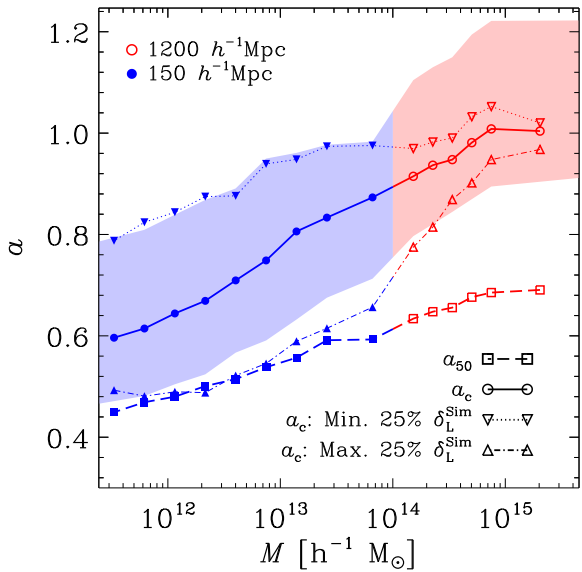}
  \caption{Mass dependence of halo collapse and formation times. Circles connected by solid
    lines show the median ``collapse redshift'', $z_{\rm c}$ (defined in eq.~(\ref{eq_collapse_time})); 
    squares connected by dashed lines show the the median half-mass formation time, $z_{50}$, at which 
    each halo's main progenitor had first assembled half of its present-day mass. Triangles 
    connected by dotted and dot-dashed lines show the medians values of $z_{\rm c}$ for haloes
    that rank in the highest and lowest quartile of $\delta_{\rm L}^{\rm Sim}$.
    Filled and open symbols are used to distinguish haloes identified in our 150 $h^{-1}$ Mpc box 
    from those in the 1200 $h^{-1}$ Mpc box, respectively. Shaded regions (shown only for $z_{\rm c}$) 
    indicate the 25th and 75th percentiles of the scatter.}
  \label{pic_freez_mass}
\end{figure}
%%%%%%%%%%%%%%%%%%%%%%%%%%%%%%%%%%%%%%%%%%%%%%%%%%%%%%%%%%%%%%%%%%%%%%%%%% 

%%%%%%%%%%%%%%%%%%%%%%%%%%%%%%%%%%%%%%%%%%%%%%%%%%%%%%%%%%%%%%%%%%%%%%%%%% 
\begin{figure}
 \centering
 \includegraphics[width=8cm,bb=0 0 735 708,keepaspectratio=true]{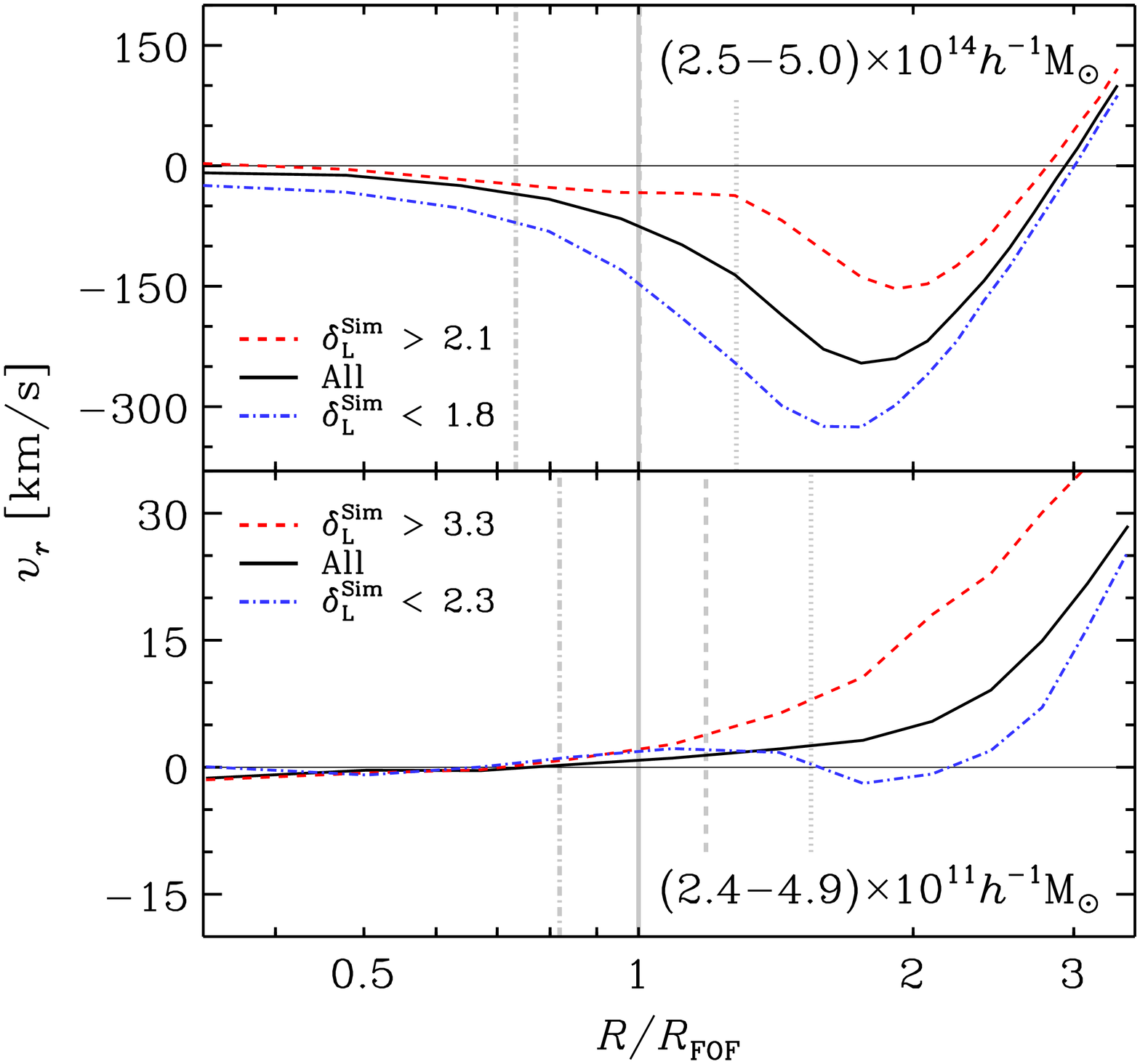}
 \caption{Solid black lines show the median radial velocity profiles of FOF haloes at 
   $z_{\rm id}=0$ in two separate mass bins. Dashed (red) and dot-dashed (blue) lines
   show the median $v_r(r)$ for haloes that rank in the highest and lowest 15 per cent
   of $\delta_{\rm L}^{\rm Sim}$. In all cases, the radial coordinate is expressed in 
   units of the mean size of the FOF haloes. Vertical grey lines indicate the typical 
   size of the haloes when identified with the SO-algorithm, using an overdensity 
   threshold of $\Delta=100$ (dotted), 200 (dashed) and 500 (dot-dashed), respectively.
   Note the different scale of the velocity axis in the upper and lower 
     panels.}
 \label{pic_radial_vel}
\end{figure}
%%%%%%%%%%%%%%%%%%%%%%%%%%%%%%%%%%%%%%%%%%%%%%%%%%%%%%%%%%%%%%%%%%%%%%%%%% 

In Figure~\ref{pic_freez_mass} we plot the mass dependence of $a_{\rm c}=(1+z_{\rm c})^{-1}$ for all $z_{\rm id}=0$ 
haloes in each simulation. 
Note that we have verified that our estimates of $z_c$ are insensitive to 
the snapshot output sequence by skipping even-numbered outputs and repeating the analysis.
Connected, filled (blue) circles show the median trend for all haloes in our
150 $h^{-1}$ Mpc box; open (red) circles correspond to haloes in our 1200 $h^{-1}$ Mpc box. 
Shaded regions in each case indicate the 25th and 75th percentiles of the scatter\footnote{
  Because many massive haloes at $z=0$ are expected to be in a state of rapid growth, the
  use of eq. (\ref{eq_collapse_time}) to estimate $z_{\rm c}$ may not be justified. In order to obtain
  a reasonable estimate of the collapse times of massive systems we
  decided to extend our 1200 $h^{-1}$Mpc box run to $z=-0.27$, which allowed us to
  track the collapse phase of even the most massive haloes identified at
  $z_{\rm id}=0$.}. As expected, 
the most massive haloes (those above a few $\times 10^{14}\, h^{-1}\, M_\odot$) have typical collapse 
times of $z_{\rm c}\approx 0$, whereas $z_{\rm c}>0$ for lower mass haloes.
The median mass dependence to $a_{\rm c}$ can be approximated by a simple linear function:
\begin{eqnarray}
 \label{eq_fit}
 a_{\rm c}=c_1\, \log_{10}(M/[h^{-1}{\rm M}_\odot])+c_0.
\end{eqnarray}
The values of the best-fitting parameters are provided in Table~\ref{tab_fit} for several 
different halo definitions.

\subsection{The collapse threshold at collapse redshift $z_{\rm c}$} \label{ss_zcoll_del}

Adopting separate collapse and identification redshifts has a subtle implication for comparing 
the predictions of the EC model to the outcome of simulations. FOF haloes defined at $z_{\rm id}=0$ are
bounded by mass (and resolution) dependent isodensity contours. Our definition of $z_{\rm c}$, however, 
defines the time at which the entire $z=0$ mass was first confined within approximately the same 
{\em physical} volume. As a result, the (comoving) overdensity of the halo at $z_{\rm c}$ will be 
lower than its $z=0$ value by a factor of $(1+z_{\rm c})^3$. To account for the different density 
contrast at collapse we multiply the radial freezing factor, $f$, used in the EC model by $(1+z_{\rm c})$. 

Additional corrections to $f$ can be made in order to account for the mass-dependence of FOF halo 
overdensities. For example, we find that haloes in our 150 $h^{-1}$Mpc box have, on average, 
$\Delta \approx 325$, whereas $\Delta \approx 270$ for those in our 1200 $h^{-1}$Mpc run 
\citep[see][for a more detailed discussion of the overdensities of FOF groups]{More2011}. 
Since $f=0.178$ sets the $z=0$ virial overdensity in the spherical 
collapse model, we modify the radial freezing factors in the EC model to match the mean 
overdensities of haloes in each of our simulations. This results in $f=0.145$ for haloes in 
our 150 $h^{-1}$Mpc box, and $f=0.155$ for those in the 1200 $h^{-1}$Mpc box.

%%%%%%%%%%%%%%%%%%%%%%%%%%%%%%%%%%%%%%%%%%%%%%%%%%%%%%%%%%%%%%%%%%%%%%%%%% 
\begin{table}
  \caption{Values obtained from fitting eq.~(\ref{eq_fit}) to the median collapse redshift 
    as a function of halo mass.}
  \begin{tabular}{lcc}\hline
    Haloes            & $c_1$ & $c_0$ \\\hline\hline
    FOF               & 0.13 & -0.88  \\
    SO ($\Delta=100$) & 0.13 & -0.81  \\ 
    SO ($\Delta=200$) & 0.12 & -0.79  \\ 
    SO ($\Delta=500$) & 0.11 & -0.66  \\\hline 
  \end{tabular}
  \label{tab_fit}
\end{table}
%%%%%%%%%%%%%%%%%%%%%%%%%%%%%%%%%%%%%%%%%%%%%%%%%%%%%%%%%%%%%%%%%%%%%%%%%% 

%%%%%%%%%%%%%%%%%%%%%%%%%%%%%%%%%%%%%%%%%%%%%%%%%%%%%%%%%%%%%%%%%%%%%%%%%% 
\begin{figure*}
  \begin{center}
    \includegraphics[width=0.95\textwidth,bb=0 0 1700 587,keepaspectratio=true]{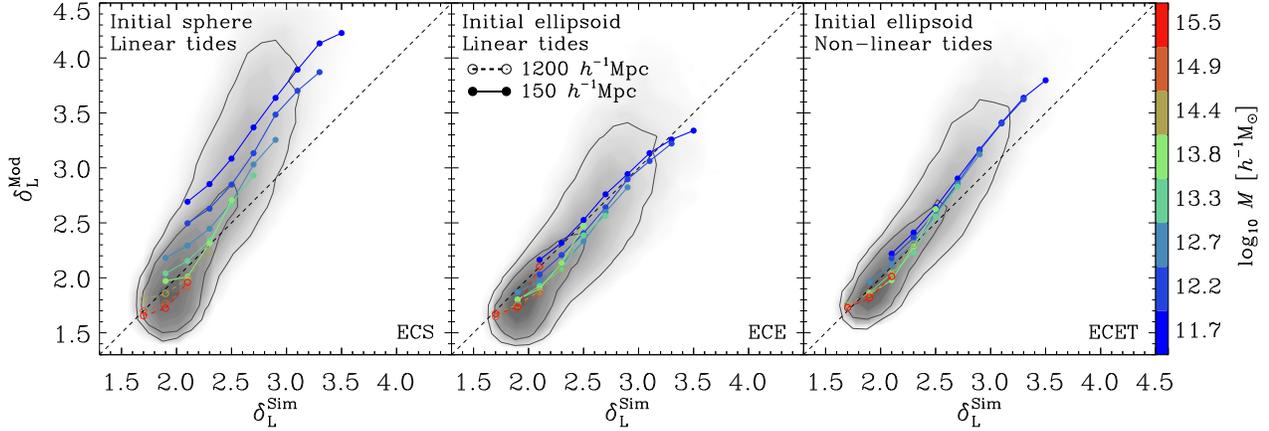}
  \end{center}
  \caption{Predicted linear overdensities of dark-matter protohaloes calculated from the 
    EC model plotted versus their measured
    overdensities. Predicted overdensities are calculated for collapse at $z_{\rm c}>0$ 
    ($z_{\rm c}$ is the collapse redshift defined by eq.~(\ref{eq_collapse_time}),
    and is explicitly calculated for each individual halo). The left-hand panel corresponds 
    to the predictions of the \citetalias[][ECS]{Bond1996} model, which assumes that each 
    protohalo occupies a spherical Lagrangian volume, and that external tidal forces evolve
    from their initial values according to linear theory. Predictions in the middle panel 
    explicitly account for the non-spherical shapes of dark-matter protohaloes, as measured in 
    the initial conditions of our simulations, but retains the linear evolution of their
    tidal fields (this model is referred to in the text as the ECE model). Finally, the
    right-most panel shows the predictions of the EC model after fully
    accounting for the triaxial shapes of dark-matter protohaloes, as well as the evolution
    of their non-linear external tidal fields (referred to as ECET). As in similar figures,
    the connected symbols show, for various mass bins, the median values of 
    $\delta_{\rm L}^{\rm Mod}$ in bins of $\delta_{\rm L}^{\rm Sim}$.}
  \label{pic_dd_all_freez}
\end{figure*}
%%%%%%%%%%%%%%%%%%%%%%%%%%%%%%%%%%%%%%%%%%%%%%%%%%%%%%%%%%%%%%%%%%%%%%%%%% 

Based on the results presented in Figure~\ref{pic_freez_mass}, the vast majority of dark-matter haloes
(apart from the most massive ones) are expected to have reached stable configurations at $z_{\rm c}\simgt 0$. 
This implies that most haloes, at the moment they are identified, are experiencing  little, if any, 
net mass accretion. We examine this point further in Figure~\ref{pic_radial_vel}, where we plot 
the average radial-velocity profiles measured in spherical bins surrounding each $z_{\rm id}=0$ 
dark-matter halo. Note that the radial coordinates have been scaled to the characteristic radius 
$R_{\rm FOF}=(3\, V_{\rm FOF}(z=0)/4\, \pi)^{1/3}$; median values of the radii
enclosing fixed overdensities of $\Delta=100$, 200, and 500 are also shown as solid, dashed and 
dot-dashed vertical lines, respectively. Panels correspond to two separate mass bins: 
$(2.5-5.0)\times 10^{14}\, h^{-1} M_\odot$ (top) and $(2.4-4.9)\times 10^{11}\, h^{-1} M_\odot$ 
(bottom). Within each mass bin separate curves show the median $v_r(r)$ profiles for all haloes 
(solid curve) as well as for the upper and lower-most 15 per cent of the $\delta_{\rm L}^{\rm Sim}$ 
distribution. 

Independent of their initial overdensity, the majority of massive haloes exhibit a strong pattern of 
infall in the regions surrounding the halo. The radius at which infall becomes substantial, however,
is a function of $\delta_{\rm L}$ (and hence, $z_{\rm c}$). This suggests,
that these systems are still accreting, and have not yet reached their final quasi-equilibrium 
state. Lower-mass haloes, on the other hand, display very little infall and are therefore not 
accreting mass \citep[see also][]{Prada2006}.

%%%%%%%%%%%%%%%%%%%%%%%%%%%%%%%%%%%%%%%%%%%%%%%%%%%%%%%%%%%%%%%%%%%%%%%%%% 
\begin{figure*}
  \centering
  \includegraphics[width=0.69\textwidth,bb=0 0 1081 1191,keepaspectratio=true]{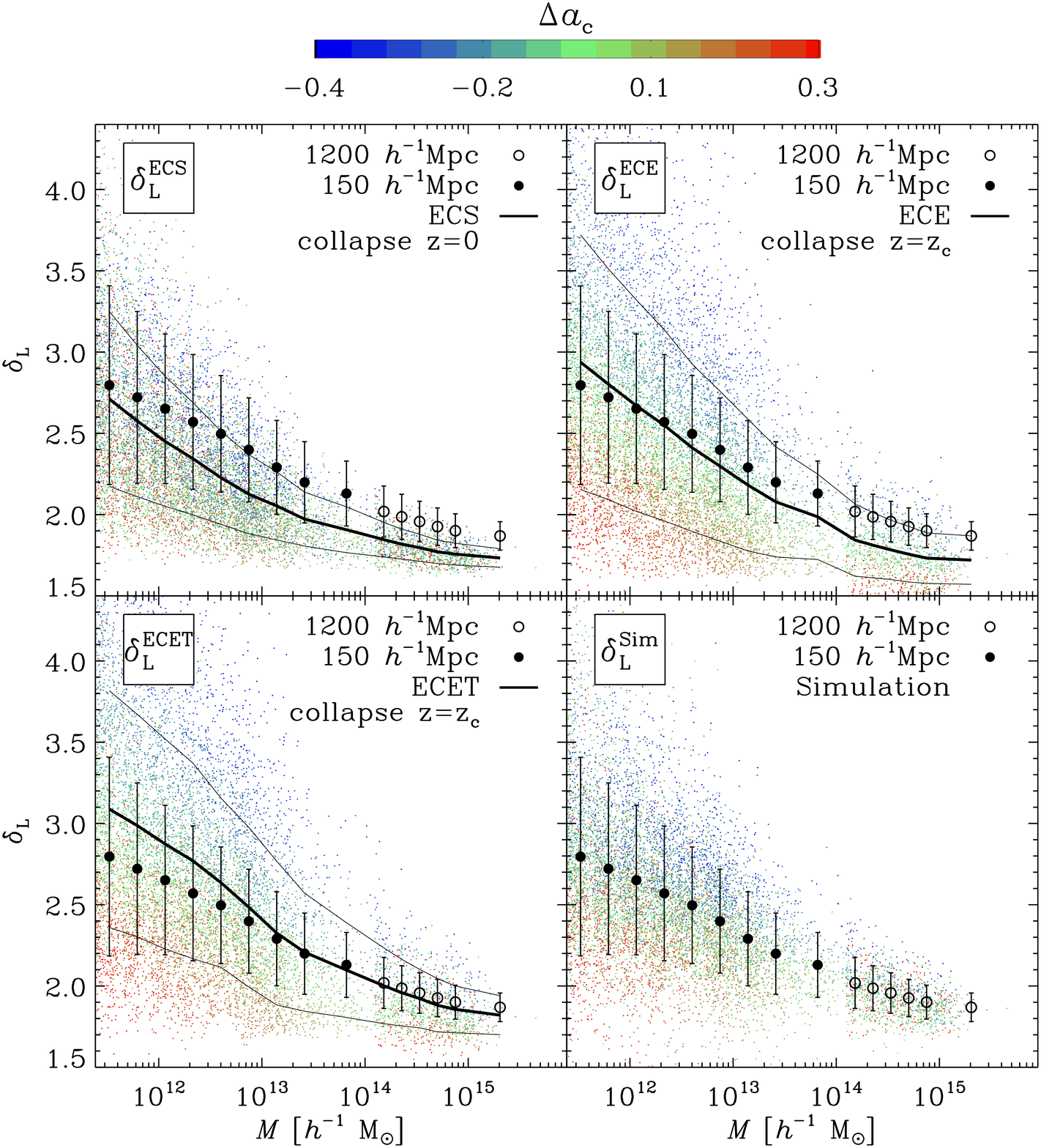}
  \caption{Linear density contrast as a function of mass. Points show the predicted value of the ECS 
    model in the upper left panel, the ECE model including collapse times in the right upper panel
    and the ECET model including collapse times in the lower left panel. The lower right panel shows 
    the values measured directly in the simulation. The points are colored according to $\Delta\, a_c$,
      the offset between the expansion factor at collapse and the corresponding mean value for haloes of the same mass
      (note that this color coding is different from what is used in Figure 1 of Paper I).
    Heavy points with error bars show the mean and standard deviation of the \emph{measured} density 
    contrast in bins of mass. Heavy lines show the mean of the predicted density contrast in each 
    panel, while thin lines indicate the standard deviation. For clarity, we have randomly
    down-sampled points in mass bins with $\simgt$1000 haloes.}
  \label{pic_scatter_delta}
\end{figure*}
%%%%%%%%%%%%%%%%%%%%%%%%%%%%%%%%%%%%%%%%%%%%%%%%%%%%%%%%%%%%%%%%%%%%%%%%%% 

The physics that determines the collapse redshift of a particular halo, or the moment at which
infall is suppressed, is not well understood. None the less, practical estimates of $z_{\rm c}$, such 
as that defined by eq.~(\ref{eq_collapse_time}), can be used to rescale the collapse redshifts
in the EC model in order to make a more meaningful prediction of the density threshold
required for the perturbation to collapse. We plot the predicted density contrast for collapse at $z_{\rm c}$, linearly 
extrapolated to $z_{\rm id}=0$, versus the measured protohalo overdensities (also linearly extrapolated 
to $z=0$) in Figure~\ref{pic_dd_all_freez}. For completeness, we include the results for all variants 
of the ellipsoidal model that we have considered, and show the median trends in the same halo-mass 
bins as before. In all cases, correcting for $z_{\rm c}>0$ significantly improves the correlation between 
the predicted a measured protohalo overdensities.

Note that the mass dependence of the $\delta-\delta$ relation predicted by the ECS model (seen already in 
Figure~\ref{pic_dd_all_today}) remains after correcting for $z_{\rm c}$. This is a result of the fact that 
$z_{\rm c}$ increases with decreasing mass, and therefore corrects the predicted overdensities of low-mass 
haloes more than those of massive ones. Haloes of $\sim 10^{12} \, h^{-1}\, M_\odot$, for example, have 
$\langle z_{\rm c}\rangle\sim 0.58$, rather than $z_{\rm c}=0$. Within the ECS model, this shift in collapse 
time is achieved by enhancing the linear density contrast of the perturbation by roughly 35 per cent. Massive 
haloes ($\simgt 10^{14}\, h^{-1}\, M_\odot$), on the other hand, have $z_{\rm c}\sim 0$ and therefore remain 
unchanged in this $\delta-\delta$ plot.

Note also that the overdensities predicted by the ECE and ECET models now follow closely the 
one-to-one line (shown in each panel as a dashed line), with the median trend differing by, at
most, 10 per cent for all halo masses and the full range of linear overdensities. The median relations 
for these models are also independent of halo mass: all individual lines in the middle and right-hand 
panels of Figure~\ref{pic_dd_all_freez} neatly overlap. The fit of a linear function, 
$\delta_{\rm L}^{\rm Mod}=A\, \delta_{\rm L}^{\rm Sim}+B$, to individual points in the $\delta-\delta$ 
plane reveals that the ECE model is consistent with a one-to-one line, while the ECS model is not.

The correction to the slope of the $\delta-\delta$ relation results from the fact that, at any
given mass scale, $\delta_{\rm L}^{\rm Sim}$ and $z_{\rm c}$ are strongly correlated. The dotted and dot-dashed 
lines in Figure \ref{pic_freez_mass} make this point clear. These curves highlight the median 
mass-dependence of $z_{\rm c}$ for haloes that, in each mass bin, rank in the highest and lowest 25 per 
cent of $\delta_{\rm L}^{\rm Sim}$. Clearly haloes that are initially denser tend to collapse earlier, 
resulting in a larger correction to their model-predicted overdensities.

\section{Discussion}\label{ch_discussion}
\subsection{The mass dependence and scatter of $\delta_{\rm L}$} \label{ss_scatter}
The dependence of $z_{\rm c}$ on $\delta_{\rm L}$ can also be seen in the lower-right panel of Figure 
\ref{pic_scatter_delta}. Here we plot the mass-dependence of $\delta_{\rm L}$ for all haloes in 
both of our simulations, and color points by $\Delta a_{\rm c}$, defined by the offset between each 
halo’s collapse time, $a_{\rm c}$ , and the median value for haloes of the same mass. The mean trends 
are shown using solid and open points for haloes in our 150 $h^{-1}$Mpc box and our 1200 $h^{-1}$Mpc box, 
respectively; the error bars indicate standard deviation.

For comparison, we also plot the {\em predicted} $\delta_{\rm L}(M)$ relations in the other three
panels, adopting the same color-coding for each. Data in the upper-left panel show the mass-dependence
of protohalo overdensities predicted by the ellipsoidal model of \citetalias{Bond1996}, tuned for $z=0$; the upper-right and
lower-left panels correspond to the ECE and ECET-model predictions (both tuned for collapse at $z_{\rm c}$).
In all cases, the mean trends at fixed mass are shown using thick solid lines, with thinner lines indicating 
the standard deviation at fixed $M$. These trends can be compared to the points with error bars, which 
reproduce the measured $\delta_{\rm L}(M)$ relation plotted in the lower-right panel. 

Note that, while the standard ECS model systematically under-predicts the linear collapse threshold,
the ECE and ECET models fare much better. Both models reproduce the mass and collapse time-dependence of $\delta_{\rm L}$ 
rather well; the average trends deviate by at most $\sim$10 per cent over roughly five orders of magnitude in mass. This is
remarkable given the simplicity of the ellipsoidal model, which overlooks entirely the complex
hierarchical growth of dark-matter haloes. Note also that these models predict a scatter in $\delta_{\rm L}$
that decreases with increasing mass, which is qualitatively consistent with the variance of
measured protohalo overdensities. However, all models typically over-predict the scatter in 
$\delta_{\rm L}$ at a given mass, and it is unclear whether different estimates of halo collapse
times will perform better in this regard.

%%%%%%%%%%%%%%%%%%%%%%%%%%%%%%%%%%%%%%%%%%%%%%%%%%%%%%%%%%%%%%%%%%%%%%%%%% 
\begin{figure}
  \centering
  \includegraphics[width=0.5\textwidth,bb=0 0 566 566,keepaspectratio=true]{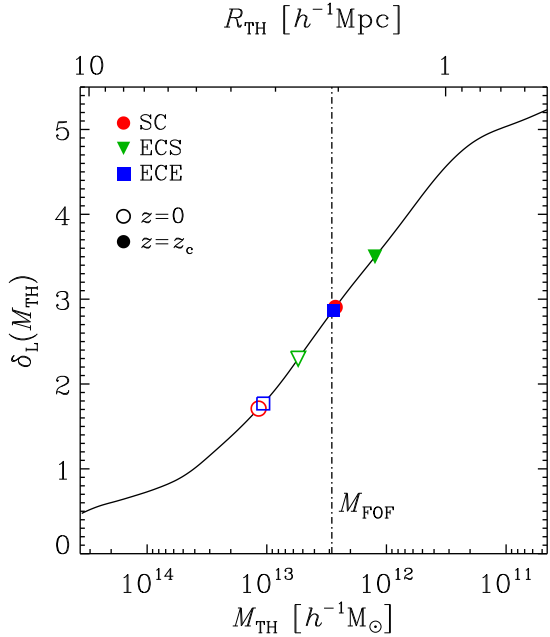}
  \caption{Excursion-set trajectory associated with the protohalo shown in Fig.~\ref{ex_freez_mass}.
    $\delta_{\rm L}(M_{\rm TH})$ is the linear density contrast extrapolated to $z=0$ after averaging within a sphere 
    of radius $R_{\rm TH}$ (corresponding to the  mass scale $M_{\rm TH}=4\pi \rho_{\rm crit}\, \Omega_{\rm M}\, R_{\rm TH}^3/3$) 
    extending around the protohalo center of mass. Open symbols highlight the points at which the 
    trajectory crosses the threshold for halo formation at $z=z_{\rm  id}=0$ evaluated using different 
    collapse models. In all cases, the EPS theory substantially overestimates the halo mass measured in 
    the simulation at  $z_{\rm id}$ (indicated by the vertical dot-dashed line). On the other hand, when
    the SC and ECE models are tuned for collapse at $z_{\rm c}$ (solid symbols), the predicted halo masses 
    are is in excellent agreement with the numerical results.}
  \label{pic_eps_example}
\end{figure}
%%%%%%%%%%%%%%%%%%%%%%%%%%%%%%%%%%%%%%%%%%%%%%%%%%%%%%%%%%%%%%%%%%%%%%%%%% 

\subsection{Implications for the extended Press-Schechter formalism} \label{ss_eps}

Our results raise questions concerning the validity of the extended Press-Schechter 
(EPS) formalism \citep{Bond1991}, particularly for low-mass haloes identified 
at late times (i.e. those with $\sigma(M)\gg \delta_{\rm c}$). In the EPS
theory, the outer boundary of a protohalo coincides  with the initial
location of a spherical mass shell that should collapse at  $z_{\rm
id}$. However, our results demonstrate that most haloes stop accreting
matter at $z_{\rm c}>z_{\rm id}$ and are essentially immutable
thereafter. For redshifts $z<z_{\rm c}$, the use of halo finding
algorithms based on density contrasts results in the (small)
pseudo-growth of the halo mass due to the decreasing background density.

We present an illustrative (and typical) example in
Figure~\ref{pic_eps_example}, where we plot the excursion-set trajectory
at the Lagrangian location of the same halo used in
Figure~\ref{ex_freez_mass}. The solid line shows the linear density
contrast (extrapolated to $z=0$) smoothed with a spherical top-hat
filter of radius $R_{\rm TH}$ and mass $M_{\rm TH}$. The open symbols indicate the
points where the trajectory crosses the threshold for collapse at
$z=z_{\rm id}=0$ computed using the SC (circle), ECS (triangle) and ECE
(square) models. In all cases, the EPS model substantially overestimates
the halo mass measured in the simulation at $z_{\rm id}$ (dot-dashed line).
This is because, contrary to the model assumption, no mass shells were
accreted onto the halo after $z_{\rm c}=1.2$. On the other hand, the EPS
predictions are rather accurate at $z_{\rm c}$ (cf. the solid  symbols
and the dot-dashed line), when the ECE (or SC) model is adopted to
predict the collapse threshold.
This can also be seen in Fig.~\ref{ex_freez_mass}, where the same symbols mark the collapse times 
predicted by the different models.

We therefore disagree with the interpretation given by \cite{Sheth2001}
that the mass dependence of the halo formation threshold $\delta_{\rm
c}$ is due to the fact that denser linear perturbations are necessary to
overcome stronger tides in order to guarantee collapse at $z_{\rm id}$.
Rather, we attribute the mass scaling to the fact that, on average,
low-mass haloes collapse and stop accreting at higher redshifts than
haloes of higher mass. Future work will focus on understanding the
physical mechanisms that prevent the collapse of the outer material
shells. Non-linear tidal interactions \citep[e.g.][]{Hahn2009} and the
geometrical overlap of the outer Lagrangian boundaries of neighbouring
haloes \citepalias{LudlowPorciani2011b} likely play a key role.

Finally, we note that the phenomenon known as ``assembly bias'' \citep[see
e.g.][]{Gao2005} simply reflects the dependence of the collapse 
threshold on $z_{\rm c}$ at fixed halo mass and identification redshift:
``old'' haloes ($z_{\rm c}\gg z_{\rm id}$) are more biased tracers of the
underlying matter distribution than ``young'' haloes  ($z_{\rm c}\simeq
z_{\rm id}$).

\section{Summary} \label{ch_summary}

We used two high-resolution simulations of structure formation in the $\Lambda$CDM 
cosmology to test how well the EC model describes the linear density contrasts
in regions that collapse to form haloes identified at $z_{\rm id}=0$. Our analysis focused on EC models of 
increasing complexity. The first (ECS) assumes that each protohalo can be approximated by a spherical 
Lagrangian tophat perturbation acted upon by linearly evolving external tidal forces, as described in 
\citetalias{Bond1996}. The second model (ECE) allows for initially non-spherical perturbations, but 
retains the linear evolution of external tides (this model was described in detail in 
\citetalias{LudlowPorciani2011b}). Finally, we consider a model (ECET) which accounts for both the 
non-spherical initial shape of protohaloes, as well as the fact that their external tidal fields evolve 
non-linearly. Our main results can be summarized as follows.

\begin{enumerate}

 \item The ECS model fails to describe the linear density contrasts measured at the sites
   of halo collapse in the initial conditions of our simulations. In this model, the required 
   density contrast for collapse (at fixed $z_{\rm c}$) is determined entirely by the surrounding tidal
   field, i.e. by $e$ and $p$. Because the average tidal field strength decreases with mass, the ECS 
   prediction is strongly mass-dependent, and is unable to reproduce the measured density 
   contrasts of haloes on an object-by-object basis, as evident in Figure~\ref{pic_dd_all_today}.
   For example, protohaloes with the same linear density contrast, but with final 
   masses of $10^{12}$ and $10^{14}\, h^{-1}M_\odot$, have predicted $\delta_{\rm L}^{\rm ECS}$
   values that differ systematically by $\sim \,$30 per cent.

 \item The mass dependence of the predicted protohalo overdensity disappears completely
   when their measured initial shapes are properly accounted for in the model calculation. This
   is because initially triaxial perturbations that align with the eigenvectors of their 
   external tidal field typically collapse at lower overdensities than their spherical 
   brethren when acted upon by strong 
   external tides \citepalias{LudlowPorciani2011b}. The mass dependence of protohalo shapes (in which lower mass
   haloes are systematically less spherical) therefore balances the higher initial density
   contrasts needed for low-mass systems to overcome the strong tidal forces in the ECS model. 
   Nevertheless, when tuned for collapse at $z=0$, the model still fails to predict the observed 
   range of Lagrangian overdensities measured in our simulations, succeeding only for very massive 
   haloes and those with low initial density contrasts. 

 \item In order to better understand these results, we developed an accurate method to measure
   the time evolution of the strength and orientation of the external tidal field acting upon
   a collapsing halo. This method reproduces the linear evolution of external tides for massive
   haloes, and for those forming in low density regions, but shows a clear non-linear evolution 
   for highly clustered haloes in the initial density field.
   Although non-linear tides act to inhibit the collapse of dense protohaloes (and therefore
   increase the model-predicted density contrasts for collapse at $z=0$), incorporating 
   these effects into the EC model only slightly improves the agreement between the model
   prediction the measured Lagrangian overdensities of protohaloes. 

 \item The main discrepancy between the predicted and measured protohalo overdensities
   can be accurately accounted for if one drops the assumption that haloes are collapsing 
   {\em today}. For a given tidal ellipticity, $e$, and prolaticity, $p$, the barrier height for 
   collapse at redshift $z_{\rm c}>z_{\rm id}$ predicted by the EC model is larger by factor of 
   $D(z_{\rm id})/D(z_{\rm c})>1$. We devised a simple method of calculating $z_{\rm c}$: it does not 
   depend on halo merger histories or on the growth of the main progenitor, but approximately
   estimates the earliest time at which the entire $z=0$ mass of the halo first reached a stable
   volume. Using this, we showed that a modified EC model that accounts for both 
   the triaxial nature of protohaloes as well as the collapse times of their descendants, can predict
   the Lagrangian density contrast of protohaloes in an unbiased way.

\item Although the ECE and ECET models provide a more faithful description of the density thresholds
    required for gravitational collapse, they rely heavily on input from the simulations. This 
    substantially reduces the predictability of these models, but illuminates possible avenues for 
    future progress. For example, a deeper understanding of the connection between the overdensities
    of protohaloes and their collapse times, $z_c$, would undoubtedly result in a much more powerful
    analytic model for halo formation. This would have important implications for theoretical modeling 
    of, among other things, large-scale structure, the halo mass function, and the merger trees of dark 
    matter halos. 

\end{enumerate}

Our results suggest that the Lagrangian overdensity of regions that collapse to form
haloes by $z=0$ increase toward lower halo masses \citep[see also][]{Robertson2009,Elia2012}.
Within the context of the standard model for ellipsoidal collapse (referred to here as
the ECS model) this behavior results from the fact that lower-mass haloes are subject
to stronger tidal distortion and therefore require larger initial overdensities to collapse
by a particular time \citep{Sheth2001}. Our results suggest a different interpretation. One difference
results from the much weaker dependence of $\delta_{\rm ec}$ on external tides when protohaloes
are modeled as triaxial ellipsoids rather than spheres. This substantially reduces the
model-predicted density threshold required for the collapse of low-mass (or strongly
sheared) perturbations by $z=0$. On the other hand, the vast majority of low-mass haloes
had already assembled their total $z=0$ masses at $z_{\rm c}>0$, suggesting that model barriers
tuned for collapse {\em today} are underestimating the true barrier height. In this 
interpretation, the shape of the $\delta_{\rm L}(M)$ relation simply reflects the mass-dependence
of halo collapse times: low-mass haloes collapse, on average, earlier than more massive 
ones, and therefore have higher initial density contrasts. This interpretation is supported
by the fact that, at fixed halo mass, $\delta_{\rm L}$ depends strongly on $z_{\rm c}$ but 
not on the shape or strength of the surrounding tidal field.

\section*{Acknowledgments}

MB acknowledges financial support from from the Deutsche Forschungsgemeinschaft through
the Transregio 33, ``The Dark Universe'', and ADL through the SFB (956), ``The Conditions
and Impact of Star Formation''. We wish to thank our referee, Aseem Paranjape,
for a constructive report that has improved this paper.

\bibliographystyle{mn2e} 
 \bibliography{paper}{}  

\appendix

\section{SO-halo finder}\label{app_so}

\begin{figure*}
 \centering
 \includegraphics[width=0.95\textwidth,bb=0 0 1700 587,keepaspectratio=true]{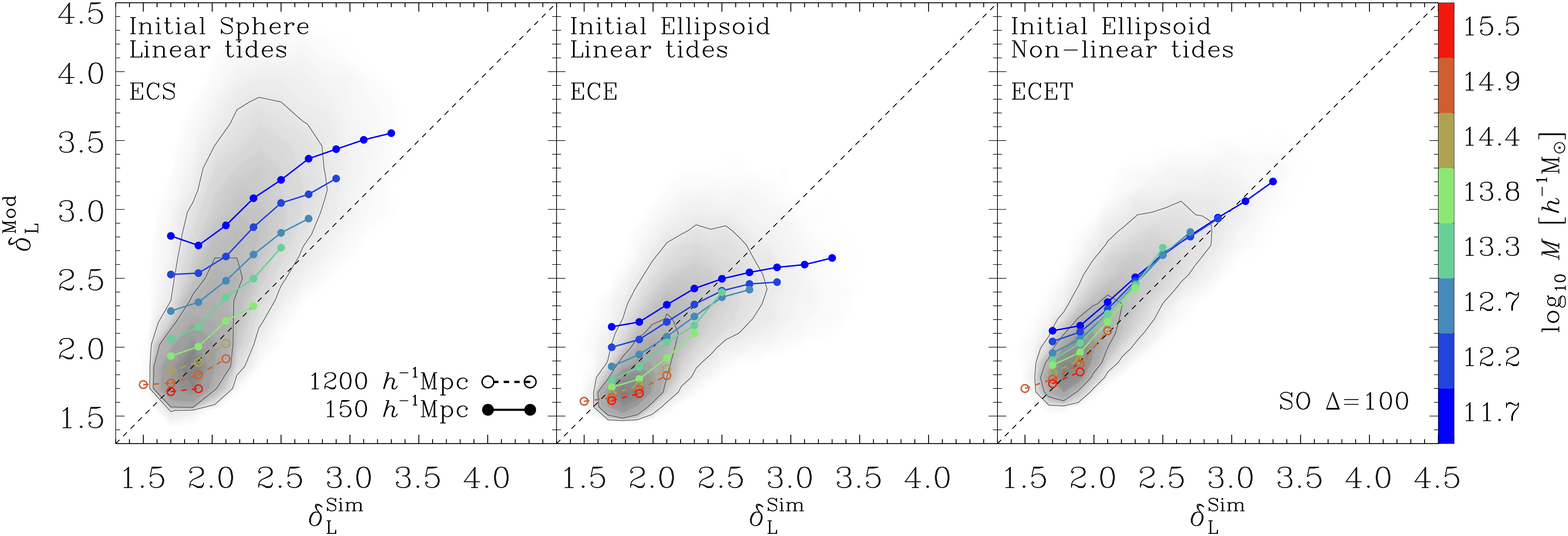}
 \includegraphics[width=0.95\textwidth,bb=0 0 1700 587,keepaspectratio=true]{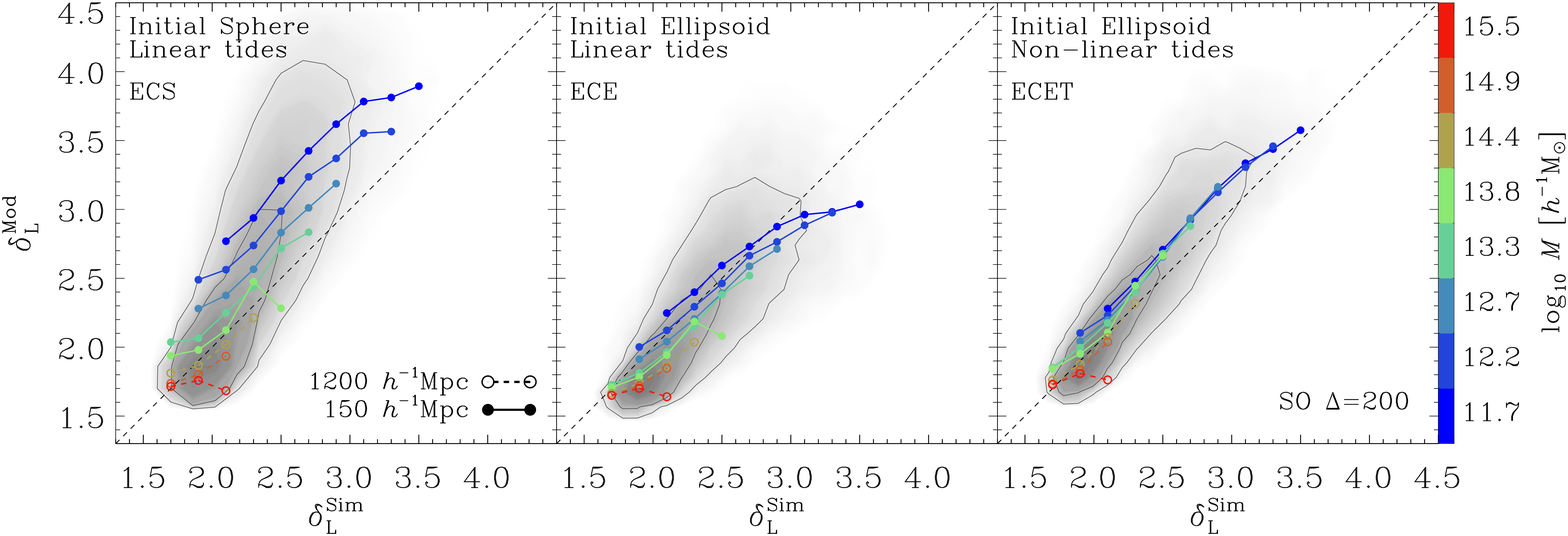}
 \includegraphics[width=0.95\textwidth,bb=0 0 1700 587,keepaspectratio=true]{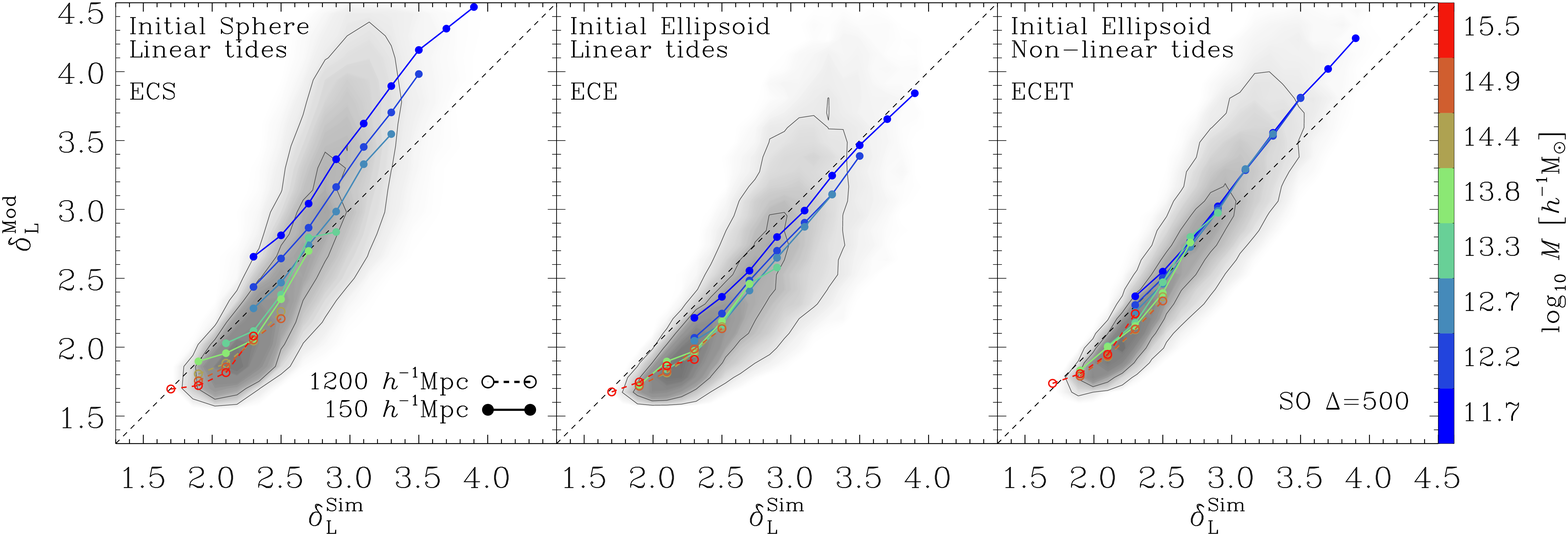}
 \caption{Same as Figure \ref{pic_dd_all_freez} but for haloes identified using the spherical overdensity 
   algorithm with density contrast thresholds of 100, 200 and 500.}
 \label{pic_dd_so}
\end{figure*}

In order to test the sensitivity of our results to our adopted (FOF) halo definition, we have repeated 
the analysis on dark-matter haloes identified using a spherical overdensity (SO) algorithm. Our SO halo finder
identifies local maxima in the evolved density field and grows spheres around them until a given density
contrast, $\Delta$, is reached. For this analysis we have adopted three different over-density values:
$\Delta=100$, 200 and 500. All aspects of the analysis were carried out as described in Section 
\ref{ch_analysis}, including modifications to the model freezing factor, $f$, required to match the 
($z=0$) SO-halo overdensities at $z=0$. 

In Figure~\ref{pic_dd_so} we plot the resulting $\delta-\delta$ (for collapse at $z_{\rm c}$ rather than at
$z=0$). In general, we find that all conclusions drawn from our analysis of FOF haloes remains valid for 
SO haloes as well, suggesting that our interpretation should not be affected by halo definition. 

\label{lastpage}

\end{document}

%% file: adsmacro.tex
%
%  These Macros are taken from the AAS TeX macro package version 5.2
%  and are compatible with the macros in the A&A document class
%  version 7.0
%  Include this file in your LaTeX source only if you are not using
%  the AAS TeX macro package or the A&A document class and need to
%  resolve the macro definitions in the TeX/BibTeX entries returned by
%  the ADS abstract service.
%
%  If you plan not to use this file to resolve the journal macros
%  rather than the whole AAS TeX macro package, you should save the
%  file as ``aas_macros.sty'' and then include it in your LaTeX paper
%  by using a construct such as:
%	\documentstyle[11pt,aas_macros]{article}
%
%  For more information on the AASTeX and A&A packages, please see:
%	http://aastex.aas.org
%       ftp://ftp.edpsciences.org/pub/aa/readme.html
%  For more information about ADS abstract server, please see:
%	http://adsabs.harvard.edu/ads_abstracts.html
%

% Abbreviations for journals.  The object here is to provide authors
% with convenient shorthands for the most "popular" (often-cited)
% journals; the author can use these markup tags without being concerned
% about the exact form of the journal abbreviation, or its formatting.
% It is up to the keeper of the macros to make sure the macros expand
% to the proper text.  If macro package writers agree to all use the
% same TeX command name, authors only have to remember one thing, and
% the style file will take care of editorial preferences.  This also
% applies when a single journal decides to revamp its abbreviating
% scheme, as happened with the ApJ (Abt 1991).

%\let\jnl@style=\rm
\def\refads@jnl#1{{\rm#1}}

\def\aj{\refads@jnl{AJ}}                   % Astronomical Journal
\def\actaa{\refads@jnl{Acta Astron.}}      % Acta Astronomica
\def\araa{\refads@jnl{ARA\&A}}             % Annual Review of Astron and Astrophys
\def\apj{\refads@jnl{ApJ}}                 % Astrophysical Journal
\def\apjl{\refads@jnl{ApJ}}                % Astrophysical Journal, Letters
\def\apjs{\refads@jnl{ApJS}}               % Astrophysical Journal, Supplement
\def\ao{\refads@jnl{Appl.~Opt.}}           % Applied Optics
\def\apss{\refads@jnl{Ap\&SS}}             % Astrophysics and Space Science
\def\aap{\refads@jnl{A\&A}}                % Astronomy and Astrophysics
\def\aapr{\refads@jnl{A\&A~Rev.}}          % Astronomy and Astrophysics Reviews
\def\aaps{\refads@jnl{A\&AS}}              % Astronomy and Astrophysics, Supplement
\def\azh{\refads@jnl{AZh}}                 % Astronomicheskii Zhurnal
\def\baas{\refads@jnl{BAAS}}               % Bulletin of the AAS
\def\bac{\refads@jnl{Bull. astr. Inst. Czechosl.}}
                % Bulletin of the Astronomical Institutes of Czechoslovakia 
\def\caa{\refads@jnl{Chinese Astron. Astrophys.}}
                % Chinese Astronomy and Astrophysics
\def\cjaa{\refads@jnl{Chinese J. Astron. Astrophys.}}
                % Chinese Journal of Astronomy and Astrophysics
\def\icarus{\refads@jnl{Icarus}}           % Icarus
\def\jcap{\refads@jnl{J. Cosmology Astropart. Phys.}}
                % Journal of Cosmology and Astroparticle Physics
\def\jrasc{\refads@jnl{JRASC}}             % Journal of the RAS of Canada
\def\memras{\refads@jnl{MmRAS}}            % Memoirs of the RAS
\def\mnras{\refads@jnl{MNRAS}}             % Monthly Notices of the RAS
\def\na{\refads@jnl{New A}}                % New Astronomy
\def\nar{\refads@jnl{New A Rev.}}          % New Astronomy Review
\def\pra{\refads@jnl{Phys.~Rev.~A}}        % Physical Review A: General Physics
\def\prb{\refads@jnl{Phys.~Rev.~B}}        % Physical Review B: Solid State
\def\prc{\refads@jnl{Phys.~Rev.~C}}        % Physical Review C
\def\prd{\refads@jnl{Phys.~Rev.~D}}        % Physical Review D
\def\pre{\refads@jnl{Phys.~Rev.~E}}        % Physical Review E
\def\prl{\refads@jnl{Phys.~Rev.~Lett.}}    % Physical Review Letters
\def\pasa{\refads@jnl{PASA}}               % Publications of the Astron. Soc. of Australia
\def\pasp{\refads@jnl{PASP}}               % Publications of the ASP
\def\pasj{\refads@jnl{PASJ}}               % Publications of the ASJ
\def\rmxaa{\refads@jnl{Rev. Mexicana Astron. Astrofis.}}%
                % Revista Mexicana de Astronomia y Astrofisica
\def\qjras{\refads@jnl{QJRAS}}             % Quarterly Journal of the RAS
\def\skytel{\refads@jnl{S\&T}}             % Sky and Telescope
\def\solphys{\refads@jnl{Sol.~Phys.}}      % Solar Physics
\def\sovast{\refads@jnl{Soviet~Ast.}}      % Soviet Astronomy
\def\ssr{\refads@jnl{Space~Sci.~Rev.}}     % Space Science Reviews
\def\zap{\refads@jnl{ZAp}}                 % Zeitschrift fuer Astrophysik
\def\nat{\refads@jnl{Nature}}              % Nature
\def\iaucirc{\refads@jnl{IAU~Circ.}}       % IAU Cirulars
\def\aplett{\refads@jnl{Astrophys.~Lett.}} % Astrophysics Letters
\def\apspr{\refads@jnl{Astrophys.~Space~Phys.~Res.}}
                % Astrophysics Space Physics Research
\def\bain{\refads@jnl{Bull.~Astron.~Inst.~Netherlands}} 
                % Bulletin Astronomical Institute of the Netherlands
\def\fcp{\refads@jnl{Fund.~Cosmic~Phys.}}  % Fundamental Cosmic Physics
\def\gca{\refads@jnl{Geochim.~Cosmochim.~Acta}}   % Geochimica Cosmochimica Acta
\def\grl{\refads@jnl{Geophys.~Res.~Lett.}} % Geophysics Research Letters
\def\jcp{\refads@jnl{J.~Chem.~Phys.}}      % Journal of Chemical Physics
\def\jgr{\refads@jnl{J.~Geophys.~Res.}}    % Journal of Geophysics Research
\def\jqsrt{\refads@jnl{J.~Quant.~Spec.~Radiat.~Transf.}}
                % Journal of Quantitiative Spectroscopy and Radiative Transfer
\def\memsai{\refads@jnl{Mem.~Soc.~Astron.~Italiana}}
                % Mem. Societa Astronomica Italiana
\def\nphysa{\refads@jnl{Nucl.~Phys.~A}}   % Nuclear Physics A
\def\physrep{\refads@jnl{Phys.~Rep.}}   % Physics Reports
\def\physscr{\refads@jnl{Phys.~Scr}}   % Physica Scripta
\def\planss{\refads@jnl{Planet.~Space~Sci.}}   % Planetary Space Science
\def\procspie{\refads@jnl{Proc.~SPIE}}   % Proceedings of the SPIE

\let\astap=\aap
\let\apjlett=\apjl
\let\apjsupp=\apjs
\let\applopt=\ao